\documentclass[twocolumn]{aastex62}

\usepackage{graphicx,amsmath,amssymb,amstext,natbib}
\usepackage{here}
\usepackage{epsf}
\usepackage{epstopdf}
\usepackage{xspace}

\usepackage[normalem]{ulem} % for regular text; use \sout{}
\usepackage{cancel} % for equations; use \cancel{}

\usepackage{color}

\newcommand{\legac}{\mbox{LEGA-C}\xspace}

\newcommand{\um}{\ensuremath{\mu\rm{m}}\xspace}
\newcommand{\kms}{\ensuremath{\rm{km\,s}^{-1}}\xspace}
\newcommand{\alphaco}{\ensuremath{\alpha_{\rm{CO}}}\xspace}
\newcommand{\gdr}{\ensuremath{\delta_{\rm{GDR}}}\xspace}

\newcommand{\uJy}{\ensuremath{\mu\rm{Jy}}\xspace}

\newcommand{\lir}{\ensuremath{L_{\rm{IR}}}\xspace}

\newcommand{\stwo}{\ensuremath{S_\mathrm{2mm}}\xspace}

\newcommand{\Tdust}{\ensuremath{T_{\rm{dust}}}\xspace}

\newcommand{\Mstar}{\ensuremath{M_{\rm{star}}}\xspace}

\newcommand{\Mht}{\ensuremath{M_{\rm{H}_2}}\xspace}

\newcommand{\Msol}{\ensuremath{\rm{M}_\odot}\xspace}

\newcommand{\fgas}{\ensuremath{f_{\rm{gas}}}\xspace}
\newcommand{\fht}{\ensuremath{f_{\rm{H}_2}}\xspace}
\newcommand{\tdep}{\ensuremath{t_{\rm{dep}}}\xspace}
\newcommand{\lprime}{\ensuremath{\rm{L}_{\rm{CO(2-1)}}'}\xspace}

\newcommand{\reff}{\ensuremath{r_\mathrm{eff}}\xspace}

\newcommand{\oii}{[O{\scriptsize II}]\xspace}

\newcommand{\etal}{et~al.\xspace}
\newcommand{\arc}{\ensuremath{''}\xspace}

\journalinfo{Accepted to ApJ, May 4, 2018}
\shortauthors{J. Spilker, et al.}
\shorttitle{Molecular Gas in Passive Galaxies at Intermediate Redshifts}

\begin{document}

\title{\uppercase{Molecular Gas Contents and Scaling Relations for Massive Passive Galaxies\\at Intermediate Redshifts from the LEGA-C Survey}}

\correspondingauthor{Justin Spilker}
\email{spilkerj@gmail.com}

\author[0000-0003-3256-5615]{Justin Spilker}
\affiliation{Department of Astronomy, University of Texas at Austin, 2515 Speedway, Stop C1400, Austin, TX 78712, USA}

\author[0000-0001-5063-8254]{Rachel Bezanson}
\affiliation{Department of Physics and Astronomy and PITT PACC, University of Pittsburgh, Pittsburgh, PA 15260, USA}

\author[0000-0001-6371-6274]{Ivana Bari\v{s}i\'{c}}
\affiliation{Max-Planck-Institut f\"{u}r Astronomie, K\"{o}nigstuhl 17, D-69117, Heidelberg, Germany}

\author[0000-0002-5564-9873]{Eric Bell}
\affiliation{Department of Astronomy, University of Michigan, 1085 South University Avenue, Ann Arbor, MI 48109-1107, USA}

\author[0000-0003-3021-8564]{Claudia del P. Lagos}
\affiliation{International Centre for Radio Astronomy Research, M468, University of Western Australia, 35 Stirling Hwy, Crawley, WA 6009, Australia}
\affiliation{ARC Centre of Excellence for All Sky Astrophysics in 3 Dimensions (ASTRO 3D), 44 Rosehill Street Redfern, NSW 2016, Australia}

\author[0000-0003-0695-4414]{Michael Maseda}
\affiliation{Leiden Observatory, Leiden University, PO Box 9513, 2300 RA Leiden, The Netherlands}

\author[0000-0002-9330-9108]{Adam Muzzin}
\affiliation{Department of Physics and Astronomy, York University, 4700 Keele St., Toronto, Ontario, M3J 1P3, Canada}

\author[0000-0003-4196-0617]{Camilla Pacifici}
\affiliation{Astrophysics Science Division, Goddard Space Flight Center, Code 665, Greenbelt, MD 20771, USA}
\affiliation{Space Telescope Science Institute, 3700 San Martin Drive, Baltimore, MD 21218, USA}

\author[0000-0001-8823-4845]{David Sobral}
\affiliation{Physics Department, Lancaster University, Lancaster LA1 4YB, UK}

\author[0000-0001-5937-4590]{Caroline Straatman}
\affiliation{Sterrenkundig Observatorium, Universiteit Gent, Krijgslaan 281 S9, B-9000 Gent, Belgium}

\author[0000-0002-5027-0135]{Arjen van der Wel}
\affiliation{Sterrenkundig Observatorium, Universiteit Gent, Krijgslaan 281 S9, B-9000 Gent, Belgium}
\affiliation{Max-Planck-Institut f\"{u}r Astronomie, K\"{o}nigstuhl 17, D-69117, Heidelberg, Germany}

\author[0000-0002-8282-9888]{Pieter van Dokkum}
\affiliation{Astronomy Department, Yale University, New Haven, CT 06511, USA}

\author[0000-0001-6065-7483]{Benjamin Weiner}
\affiliation{Steward Observatory, University of Arizona, 933 North Cherry Ave., Tucson, AZ 85721, USA}

\author[0000-0001-7160-3632]{Katherine Whitaker}
\affiliation{Department of Physics, University of Connecticut, 2152 Hillside Road, Unit 3046, Storrs, CT 06269, USA}

\author[0000-0003-2919-7495]{Christina C. Williams}
\altaffiliation{NSF Fellow}
\affiliation{Steward Observatory, University of Arizona, 933 North Cherry Ave., Tucson, AZ 85721, USA}

\author[0000-0002-9665-0440]{Po-Feng Wu}
\affiliation{Max-Planck-Institut f\"{u}r Astronomie, K\"{o}nigstuhl 17, D-69117, Heidelberg, Germany}

%%%%%%%%%%%%%%%%%%%%%%%%%%%%%%%%%%%%%%%%%%%%%%%%%%%%%%%%%%%%%%%%%%%%%%%%%%%%%%%%%%%%%
%%%%%%%%%%%%%%%%%%%%%%%%%%%%%%%%%%%% ABSTRACT %%%%%%%%%%%%%%%%%%%%%%%%%%%%%%%%%%%%%%%
%%%%%%%%%%%%%%%%%%%%%%%%%%%%%%%%%%%%%%%%%%%%%%%%%%%%%%%%%%%%%%%%%%%%%%%%%%%%%%%%%%%%%
\begin{abstract}

A decade of study has established that the molecular gas properties of star-forming galaxies follow coherent scaling relations out to $z\sim3$, suggesting remarkable regularity of the interplay between molecular gas, star formation, and stellar growth. Passive galaxies, however, are expected to be gas-poor and therefore faint, and thus little is known about molecular gas in passive galaxies beyond the local universe. Here we present deep Atacama Large Millimeter/submillimeter Array (ALMA) observations of CO(2--1) emission in 8 massive ($\Mstar \sim 10^{11}$\,\Msol) galaxies at $z\sim0.7$ selected to lie a factor of 3--10 below the star-forming sequence at this redshift, drawn from the Large Early Galaxy Astrophysics Census (LEGA-C) survey. We significantly detect half the sample, finding molecular gas fractions $\lesssim$ 0.1. We show that the molecular and stellar rotational axes are broadly consistent, arguing that the molecular gas was not accreted after the galaxies became quiescent. We find that scaling relations extrapolated from the star-forming population over-predict both the gas fraction and gas depletion time for passive objects, suggesting the existence of either a break or large increase in scatter in these relations at low specific star formation rate. Finally, we show that the gas fractions of the passive galaxies we have observed at intermediate redshifts are naturally consistent with evolution into local massive early-type galaxies by continued low-level star formation, with no need for further gas accretion or dynamical stabilization of the gas reservoirs in the intervening 6 billion years.

\end{abstract}


%%%%%%%%%%%%%%%%%%%%%%%%%%%%%%%%%%%%%%%%%%%%%%%%%%%%%%%%%%%%%%%%%%%%%%%%%%%%%%%%%%%%%
%%%%%%%%%%%%%%%%%%%%%%%%%%%%%%%%%% Introduction %%%%%%%%%%%%%%%%%%%%%%%%%%%%%%%%%%%%%
%%%%%%%%%%%%%%%%%%%%%%%%%%%%%%%%%%%%%%%%%%%%%%%%%%%%%%%%%%%%%%%%%%%%%%%%%%%%%%%%%%%%%
\section{Introduction} \label{intro}

The processes by which galaxies grow and evolve are intimately linked to the accretion and conversion of gas into stars. In particular, because stars form from molecular gas \citep[e.g.,][]{schruba11}, the heating, cooling, and transport of gas from outside and within galaxies play a large role in determining how efficiently a galaxy can form stars, and the overall mass of stars that can be formed. By and large, galaxies form stars across cosmic time in equilibrium with the supply of fresh gas from accretion and mergers; the interplay between gas accretion, outflows, star formation, and mergers naturally regulates the growth of galaxies \citep[e.g.,][]{dave11,dave12,lilly13,peng14}. 

Surveys of increasingly large numbers of galaxies have shown that the majority of galaxies  exhibit a relatively tight and nearly-linear relationship between the current star formation rate (SFR) and the mass of stars already formed (\Mstar; e.g., \citealt{noeske07,franx08,whitaker12,whitaker14,speagle14,schreiber16}). The intrinsic scatter in the relationship is $\approx0.3$\,dex, and objects that lie near it are generally considered to be `normal' galaxies. The normalization of the `star-forming sequence' increases rapidly with redshift, implying much more rapid galaxy growth in the early universe as compared to today, or an overall increase in the SFR density function \citep{sobral14}. At the most massive end, however, an ever-larger fraction of galaxies exhibit markedly depressed SFRs (or specific SFR, sSFR\,$\equiv$\,SFR$/ \Mstar$). This transition occurs near the break in the stellar mass function, $\log \Mstar/\Msol \gtrsim 10.5-11$, and does not appear to evolve significantly with redshift (e.g., \citealt{peng10b}, though see also \citealt{gavazzi15,tomczak16}). Star formation in these massive galaxies appears to be efficiently shut off (`quenched') as they transition to the red sequence, but the physical mechanisms responsible for this quenching are still unclear.

After a decade of extensive observational investment, the relationship between gas supply and the growth of star-forming galaxies has become more clear, both at low redshift \citep[e.g.,][]{saintonge11,saintonge17,bothwell14} and in the distant universe \citep[e.g.,][]{tacconi13,magnelli14,papovich16,scoville16}. For galaxies with SFRs near and above the star-forming sequence, scaling relations have been derived relating the properties of the molecular interstellar medium (ISM) with other galaxy properties \citep[e.g.,][]{genzel15,scoville17,tacconi18}, with remarkably good agreement between various tracers of the molecular gas. In general, the molecular gas fraction $\fht \equiv \Mht/\Mstar$ increases rapidly with redshift, is elevated for galaxies well above the star-forming sequence, and shows a weak decline with increasing \Mstar. The gas depletion time, $\tdep \equiv \Mht/\mathrm{SFR}$, a measure of how long a galaxy could continue to form stars at its current rate before exhausting its gas supply, shows a much weaker evolution with redshift, is shorter for galaxies above the star-forming sequence, and is either constant or weakly increases towards high \Mstar. Typical values for Milky Way-mass star-forming galaxies at $z\sim0$ are $\fht \sim$ few\,$\times10^{-2}$  and $\tdep \sim 1$\,Gyr, and at $z\sim1$, $\fht \sim 0.5-0.8$ and $\tdep \sim 0.7$\,Gyr \citep{tacconi18}.

The recent advent of large samples of `normal' star-forming galaxies with molecular gas measurements has been particularly useful at high redshifts, providing extremely valuable reference samples for other galaxies that may be less `normal.' Of particular relevance for the buildup of the quiescent galaxy population, molecular gas observations have now also targeted smaller samples of galaxies thought to be actively quenching star formation for various reasons \citep[e.g.,][]{geach13,spilker16b,popping17} or that show spectral signatures of quenching in the past $<$1\,Gyr \citep{suess17}. These studies indicate that, while the suppression of star formation does not require the complete removal or depletion of the molecular gas, the quenching processes do appear to lower \fht at fixed mass compared to the reference samples in most cases.

Because of the relationship between \Mht and SFR, however, much less is known about the molecular ISM in galaxies well below the star-forming sequence; the expected molecular masses require very sensitive observations and substantial integration times even with the supreme sensitivity of ALMA. Observations of local massive and passive early-type galaxies, for example, reveal gas fractions an imposing 1--2 orders of magnitude lower than their star-forming counterparts \citep{young11,davis13,davis16}. Because these local galaxies show few signs of recent star formation for the past many Gyr, however, it is not clear to what extent inferences about the suppression of star formation in these galaxies also apply in the distant universe, closer to the epoch at which galaxies first became quiescent.

The main focus of our work here is to determine whether the scaling relations derived from observations of star-forming galaxies and cosmological simulations can accurately predict or be extrapolated down to passive galaxies at intermediate redshift.  We present ALMA observations of a sample of 8 massive ($\log \Mstar/\Msol > 10.8$) galaxies at $z\sim0.7$ from the \legac survey selected to lie a factor of 3--10 below the star-forming sequence at this redshift. We observed the CO(2--1) transition, a tracer of the molecular ISM. In Section~\ref{data}, we describe the parent \legac sample, our ALMA observations, and our measurements of \Mht. Section~\ref{results} provides a broad overview of our basic results, and we compare the stellar and molecular dynamics of our detected galaxies. In Section~\ref{discussion}, we compare the gas fractions and depletion times derived for our sample with observationally-based scaling relations and with the EAGLE cosmological simulation. We discuss the implications our observations have for the understanding of galaxy quenching in Section~\ref{quenching}, and conclude in Section~\ref{conclusions}. Throughout, we assume a flat $\Lambda$CDM cosmology with $\Omega_m=0.307$ and $H_0=67.7$\,\kms\,Mpc$^{-1}$ \citep{planck15}.

%%%%%%%%%%%%%%%%%%%%%%%%%%%%%%%%%%%%%%%%%%%%%%%%%%%%%%%%%%%%%%%%%%%%%%%%%%%%%%%%%%%%%
%%%%%%%%%%%%%%%%%%%%%%%%%%%%%%%%%% Observations %%%%%%%%%%%%%%%%%%%%%%%%%%%%%%%%%%%%%
%%%%%%%%%%%%%%%%%%%%%%%%%%%%%%%%%%%%%%%%%%%%%%%%%%%%%%%%%%%%%%%%%%%%%%%%%%%%%%%%%%%%%
\section{Data and Analysis} \label{data}

%%%%%%%%%%%%%%%%%%%%%%%%%%%%%%%%%%% Selection %%%%%%%%%%%%%%%%%%%%%%%%%%%%%%%%%%%%%%%
\subsection{\legac and Selection of Quiescent Galaxies} \label{selection}

We selected galaxies for CO(2--1) observations from the \legac survey of $0.6 < z < 1.0$ galaxies. The survey is described in detail by \citet{vanderwel16}. Briefly, galaxies were selected based on $K$-band magnitude from the UltraVISTA catalog described in \citet{muzzin13a}. The \legac survey consists of $\sim$3200 galaxies observed with 20 hour integrations using the VIMOS spectrograph on the Very Large Telescope. The \legac spectra yield high signal-to-noise detections of the stellar absorption features and continuum, allowing determinations of the age of the stellar populations, metallicities, and stellar velocity dispersions. Because the \legac survey targets the COSMOS extragalactic survey field, full panchromatic spectral energy distribution (SED) information is available, along with morphological information from \textit{Hubble Space Telescope} observations. The galaxies studied in this work were selected from the first data release (DR1) catalog for which VIMOS observations were completed by early 2016, consisting of 644 objects in the primary sample with good spectra. This early sample represents a subset of the full survey area and is not biased with respect to the full \legac sample in terms of, e.g., K-band magnitude. Note that, while the galaxies were selected from the initial DR1 catalog, the figures and table in this work use values from the updated DR2 catalog (1989 galaxies; Straatman \etal, in prep.). Stellar masses of the \legac sample have been measured by fitting the photometric SED using FAST \citep{kriek09} assuming a \citet{chabrier03} initial mass function. 

\subsubsection{SFR Estimation}

The primary SFR estimates for the \legac sample come from modeling of the spectral energy distribution of our target galaxies, from the ultraviolet to the mid-infrared. SFRs include the unobscured and obscured components based on UV and IR (24\,\um) fluxes. The low resolution of the \textit{Spitzer}/MIPS 24\,\um imaging requires the use of a deblending procedure to assign the measured 24\,\um flux to K-band detected galaxies \citep{muzzin13a}. In our final ALMA sample, described further below, two objects (IDs 130284 and 132776) are near bright 24\,\um sources, where the deblending is potentially unreliable.  For all objects in the ALMA sample, the inclusion of the 24\,\um fluxes increases the inferred SFR by a factor of 2.5, on average. Unsurprisingly, none of these objects with low SFR are detected in \textit{Herschel Space Observatory} imaging of the COSMOS field.

Several authors have noted that the IR luminosity (or observed-frame 24\,\um luminosity, generally the only available tracer of dust emission at low SFR and high redshift) can overestimate the obscured SFR of quiescent galaxies, in some cases quite severely \citep[e.g.,][]{salim09,hayward14,utomo14,man16}. This can be due to several effects, including the fact that the IR luminosity is a long-lived tracer compared to the instantaneous SFR, and 24\,\um emission can be boosted by dust heating unrelated to star formation, including circumstellar dust heated by intermediate-age AGB stars, extended cirrus dust heated by old stellar populations, or weak nuclear activity. The latter three result in a higher \lir/SFR ratio, or equivalently only a fraction of the observed \lir should be considered in the calculation of the SFR. 

Of these options, weak nuclear activity is unlikely, with the 24\,\um emission expected from stacking of X-ray images falling three orders of magnitude below the observed emission for intermediate-redshift quiescent galaxies \citep{fumagalli14}. Dust heating from AGB stars and cirrus dust are potentially more relevant. \citeauthor{fumagalli14} used a stacking analysis of MIPS/24\,\um images of quiescent galaxies at $0.3<z<2.5$, and found that the SFR determined from these images could be overestimated by an order of magnitude. However, there is reason to believe the situation is not so dire for our own selection. \citeauthor{fumagalli14} focused on galaxies much more quiescent than ours, 20--40$\times$ below the star-forming sequence, compared to our own selection of 3--10$\times$ below. The objects in our sample do not reach such low sSFR even if none of the IR emission is related to star formation and only the SFR based on the rest-UV is considered. The SFR overestimation becomes less severe at higher sSFR because true obscured star formation rapidly outshines the lower-level IR emission unrelated to SF. For the sSFRs typical of our sample, the work by \citet{fumagalli14} indicates that the SFRs should be overestimated by less than a factor of $\sim2$.

\subsubsection{ALMA Selection}

\begin{figure*}[htb]
\begin{centering}
\includegraphics[width=0.495\textwidth]{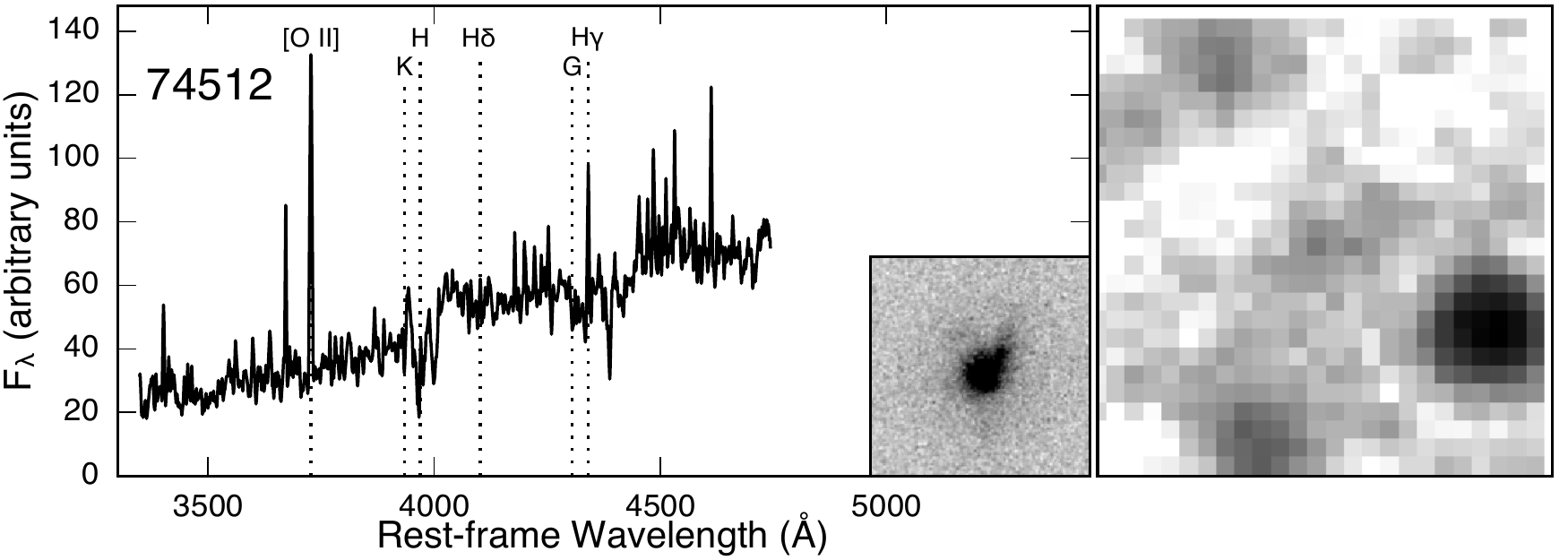}
\includegraphics[width=0.495\textwidth]{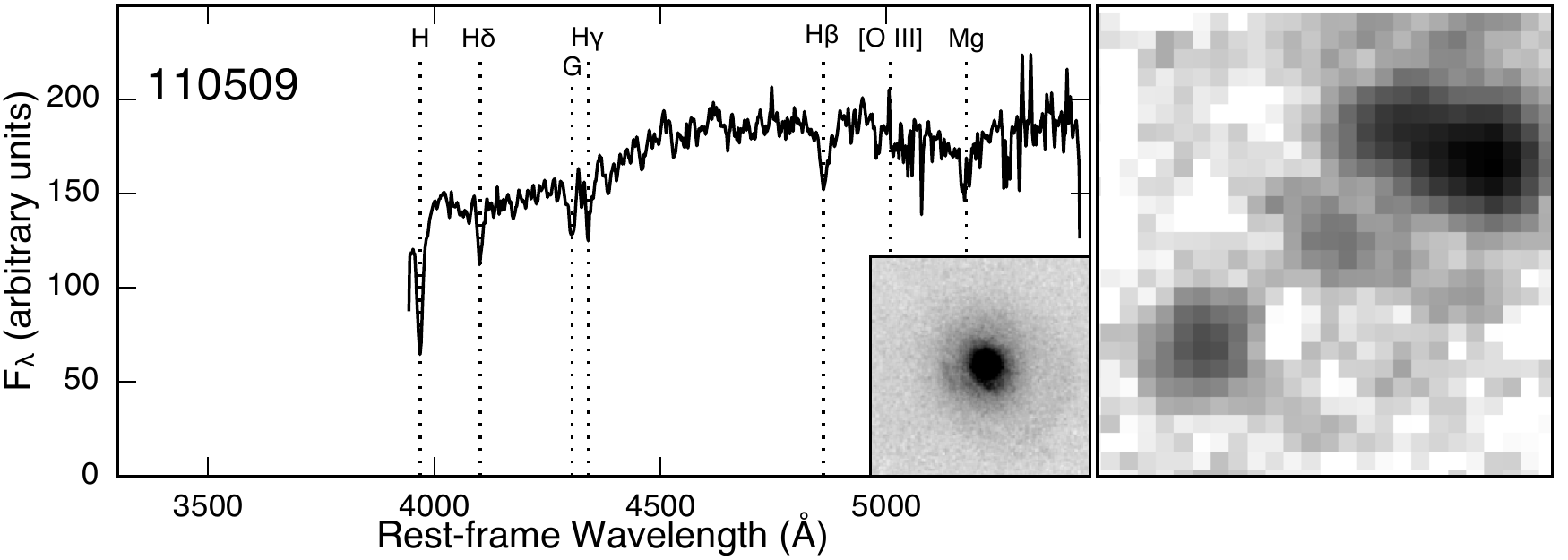}

\includegraphics[width=0.495\textwidth]{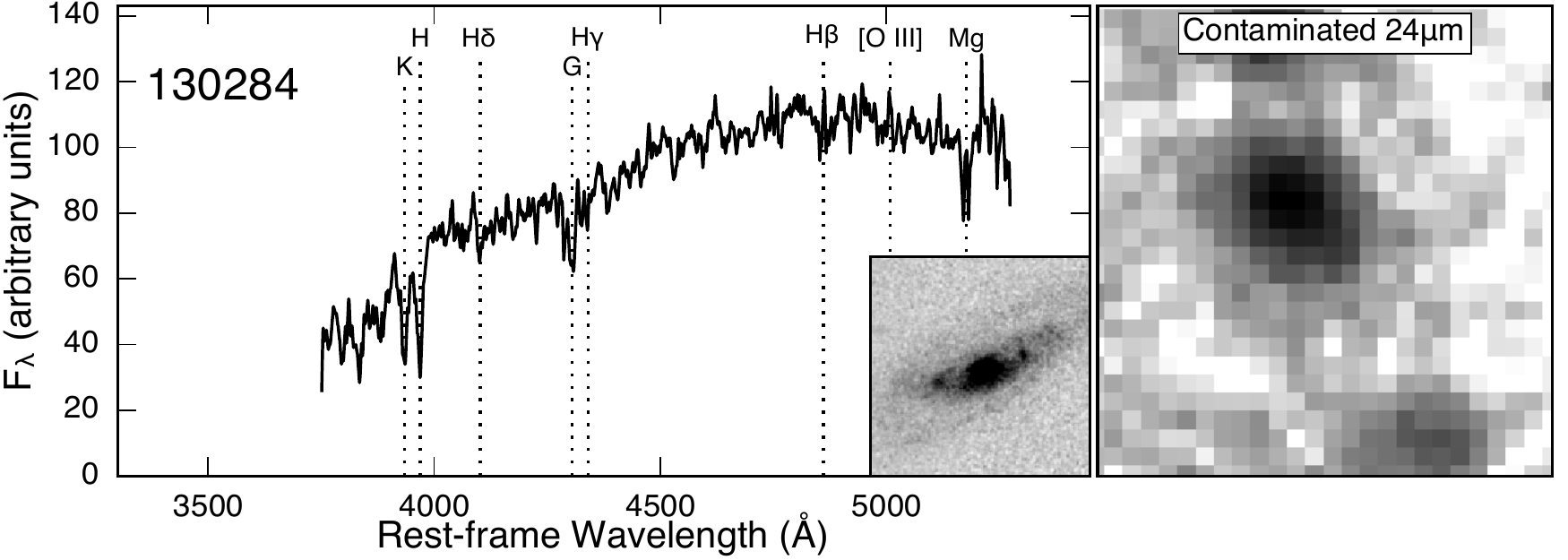}
\includegraphics[width=0.495\textwidth]{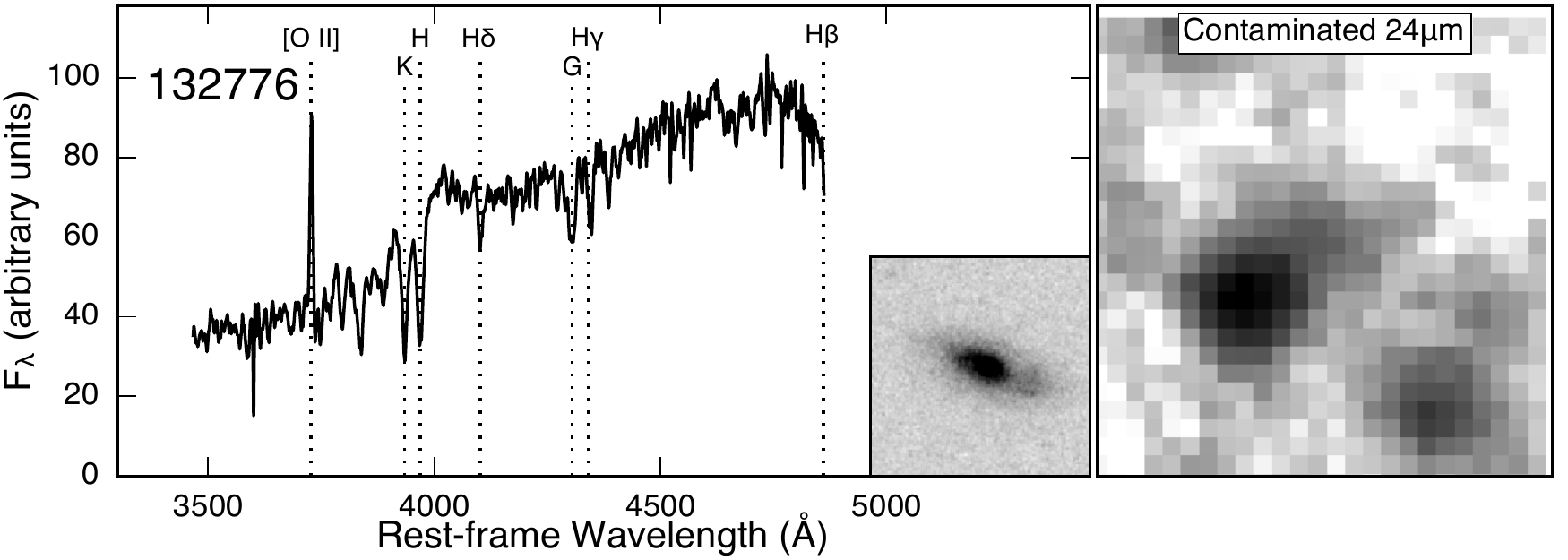}

\includegraphics[width=0.495\textwidth]{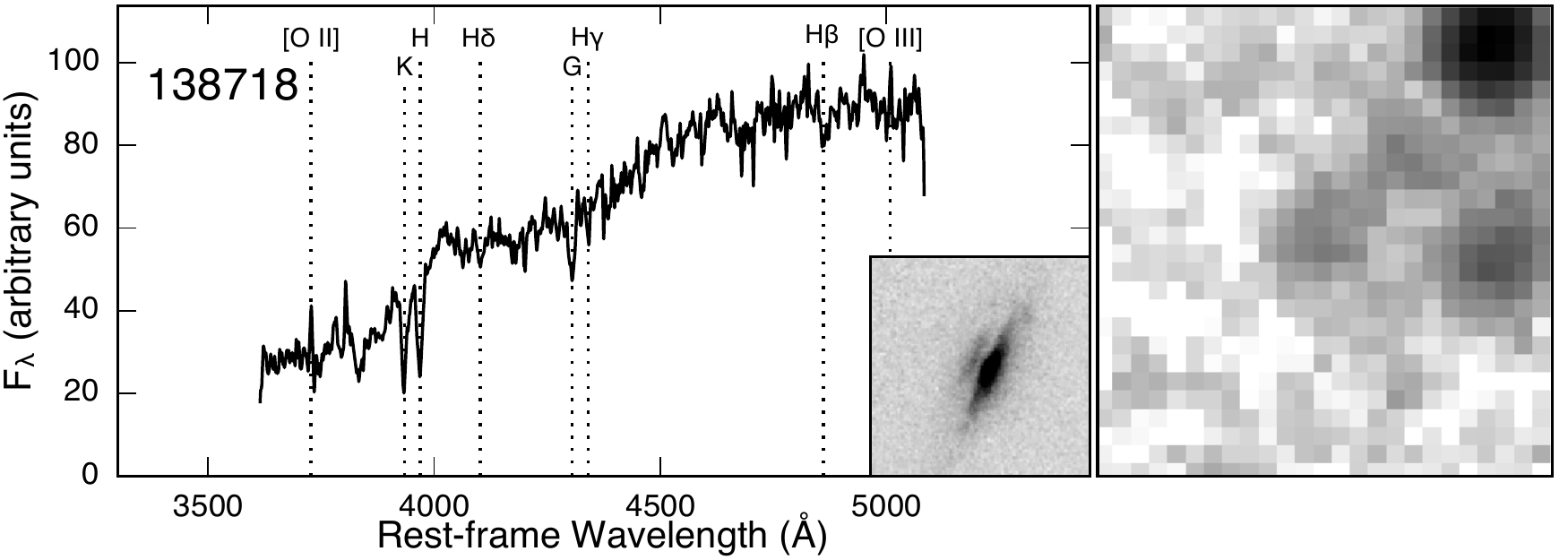}
\includegraphics[width=0.495\textwidth]{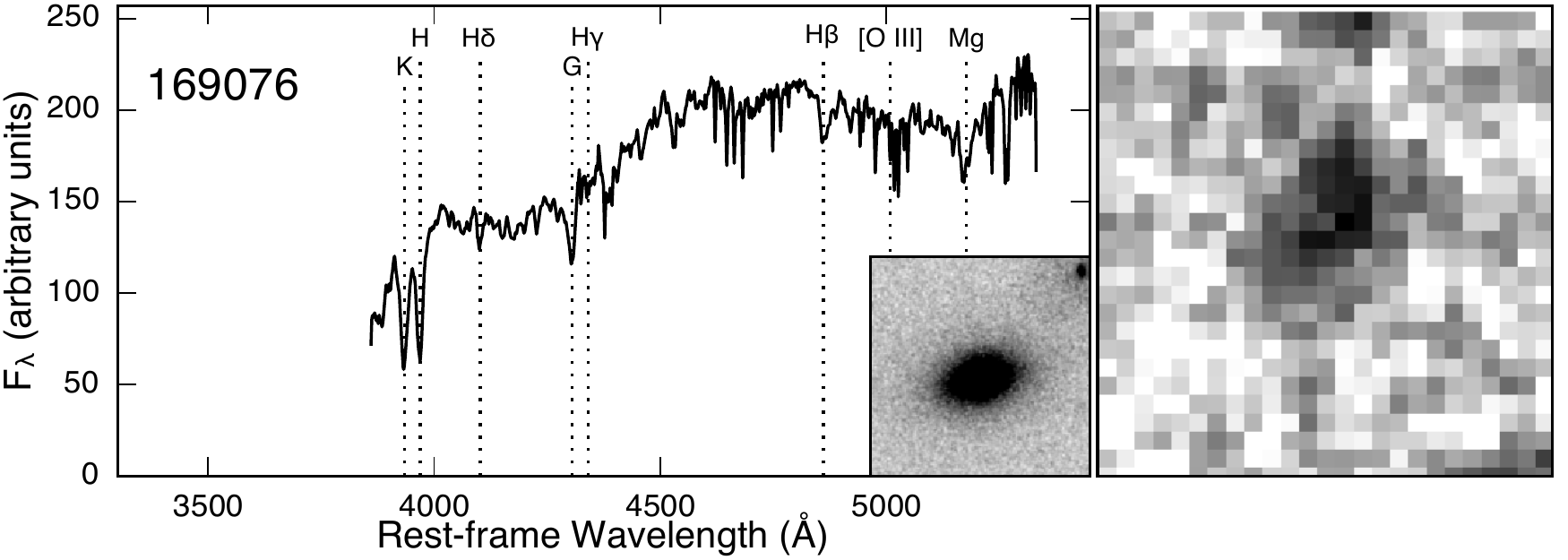}

\includegraphics[width=0.495\textwidth]{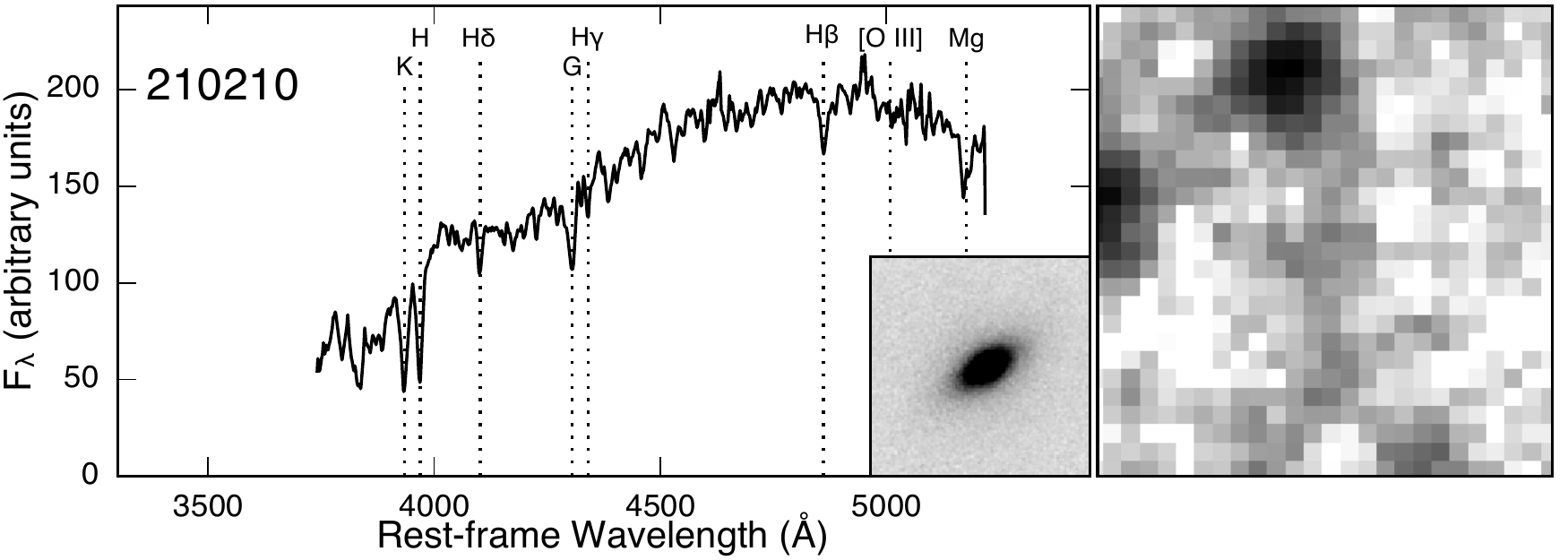}
\includegraphics[width=0.495\textwidth]{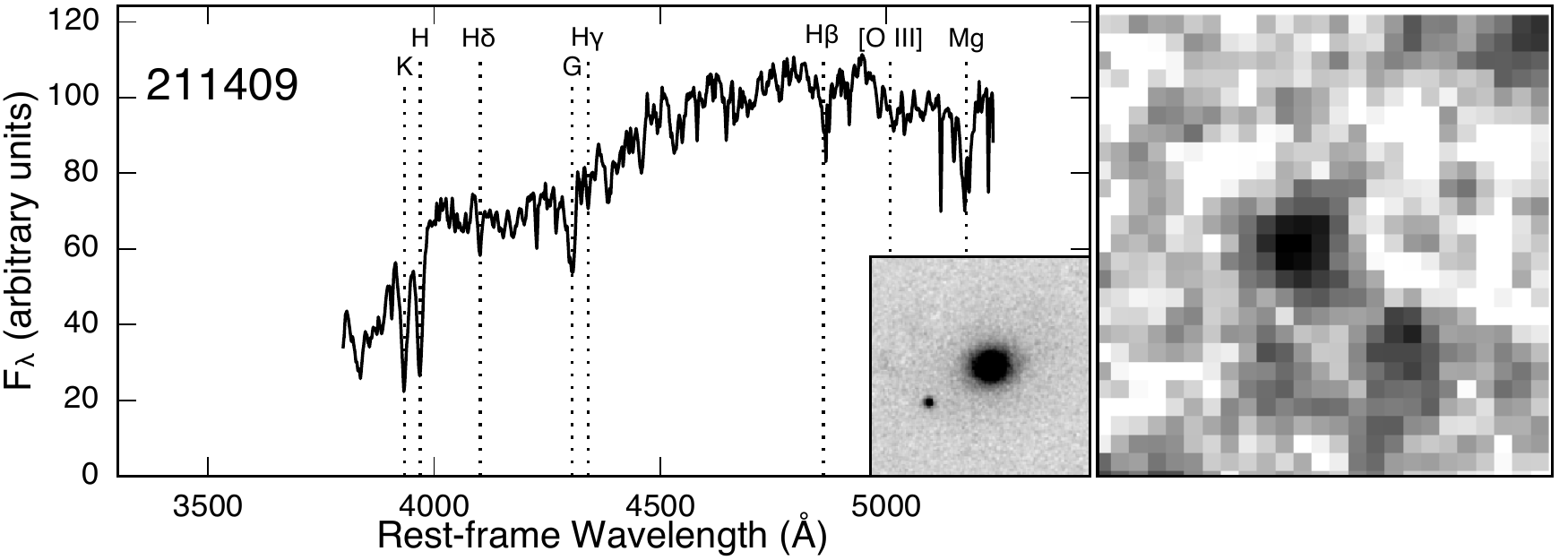}

\end{centering}
\caption{
\textit{Left:} Rest-frame optical LEGA-C spectra for each observed target, with common emission and absorption lines indicated. Spectra have been smoothed with a Gaussian of FWHM 4\,\AA. 
\textit{Insets:} \textit{HST}/ACS F814W images of each galaxy. Cutouts are 4$\times$4\arc. 
\textit{Right:} \textit{Spitzer}/MIPS 24\,\um images of each target, logarithmically scaled. Cutouts are 30$\times$30\arc. Most galaxies are clearly or marginally detected at 24\,\um, although two objects have photometry that is likely contaminated by nearby galaxies.
}
\label{fig:legacspec}
\end{figure*}

We selected galaxies from the \legac sample with spectroscopic redshift $z<0.84$, for which the CO(2--1) line is observable with the ALMA Band 4 receivers. We selected massive galaxies, with $\log \Mstar/\Msol > 10.8$ and SFR 3--10$\times$ below the $z\sim0.7$ star-forming sequence as defined by \citet{whitaker12}.  Given the intrinsic scatter in this sequence ($\approx0.3$\,dex), this selects objects with SFRs lower than $\approx$90\% of all galaxies at this redshift. We further required that SFR\,$>$\,2.5\,\Msol/yr, intended to restrict the sample to galaxies that could plausibly be detected in CO in a reasonable integration time with ALMA. This resulted in a sample of 65 galaxies, from which we selected 8 for observation with ALMA, listed in Table~\ref{tab:targets}. We aimed to include galaxies with a variety of morphologies, both obvious spiral galaxies and bulge-dominated elliptical galaxies. Two of the eight galaxies (IDs 110509 and 169076) are identified as radio-loud AGN \citep{barisic17}; both lack optical emission lines and are classified as low-excitation radio galaxies. The sensitivity of the best-available radio imaging of the COSMOS field \citep{smolcic17} is not sufficient to detect radio continuum emission related to star formation for our sample (approximate 3$\sigma$ limiting SFR$>$18\,\Msol/yr). The VIMOS spectra of our 8 galaxies are shown in Figure~\ref{fig:legacspec}.

For two objects in the ALMA sample, the LEGA-C spectra clearly detect the \oii $\lambda$3727 doublet, which can also be used as an SFR indicator, although one highly susceptible to dust extinction and low-level nuclear activity. Converting this line flux to an SFR yields values less than the UV$+$IR SFRs, as expected, but within $\sim$2$\times$ the UV-only SFR. The remaining objects show little evidence for nebular emission lines also used as SFR indicators (e.g., H$\beta$), with any emission generally overwhelmed by the strong absorption features. 

\begin{figure}[htb]
\includegraphics[width=\columnwidth]{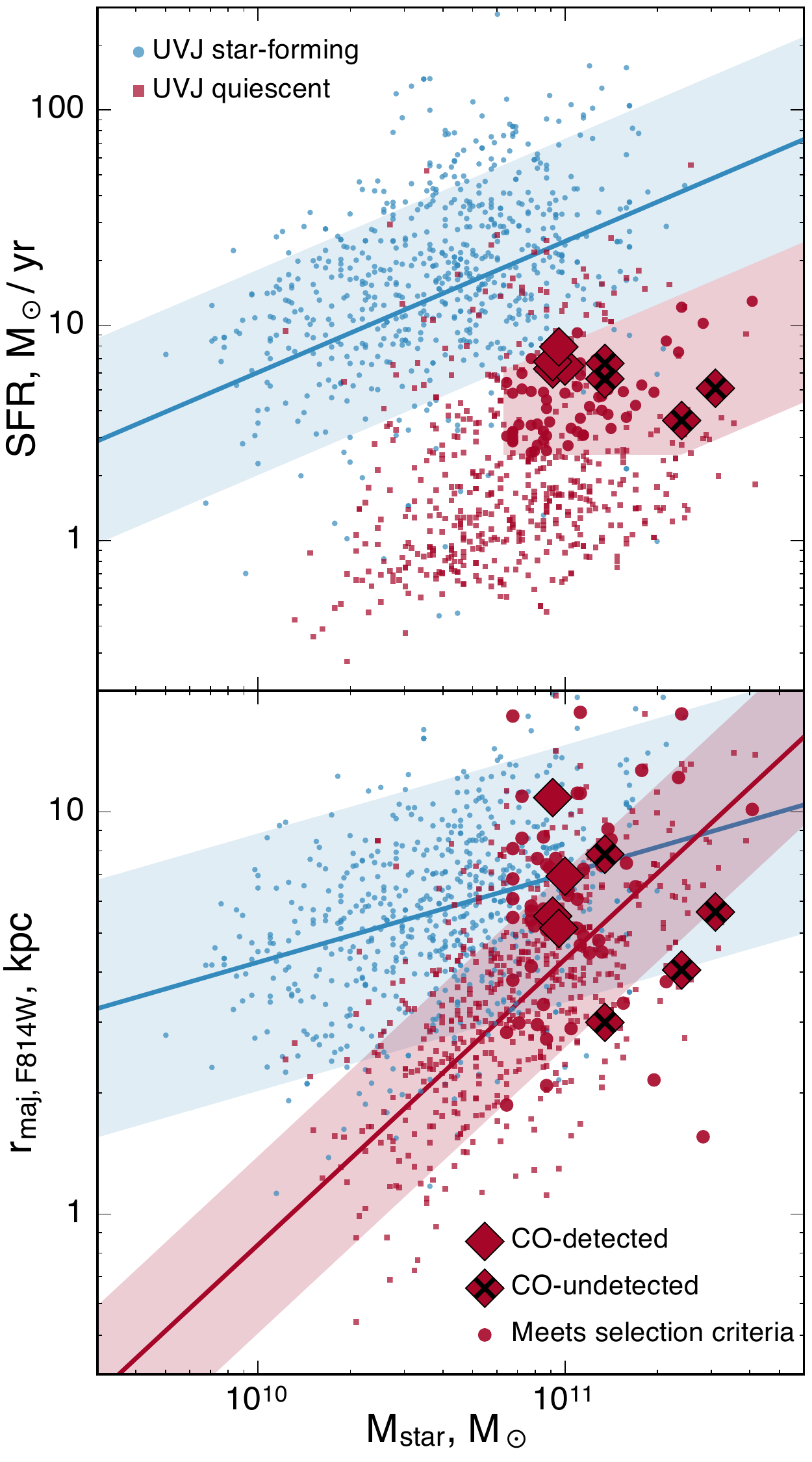}
\caption{
The selection of the ALMA-observed massive, passive sample with respect to the full \legac sample. In both panels, the \legac sample galaxies are color-coded blue (red) if they are classified as star-forming (quiescent) in rest-frame $UVJ$ space (see Figure~\ref{fig:selection2}). Red circles are all \legac galaxies that meet our selection criteria, while the ALMA-observed objects are shown with large red diamonds; CO-undetected objects are also marked with a black `$\times$'.
\textit{Top:} The ALMA sample was primarily selected based on stellar mass and SFR; see text for details. The blue line shows the star-forming sequence at $z=0.7$ from \citet{whitaker12}, and the blue shaded region encompasses SFRs a factor of 3 above and below the relation. The red shaded region shows our selection box of massive and passive galaxies.
\textit{Bottom:} Size-mass relation for the \legac and ALMA samples. Blue and red lines and regions show the size-mass relations for star-forming and quiescent galaxies at $z=0.75$ from \citet{vanderwel14}.
}
\label{fig:selection}
\end{figure}

In Figure~\ref{fig:selection} we show the selection of the ALMA sample in the SFR--\Mstar plane with respect to the full \legac sample and all other \legac objects that meet our selection criteria. We also show the size-mass relation for star-forming and quiescent galaxies, with the best-fit relations determined by \citet{vanderwel14}. Because we chose objects with a range of morphologies, it is unsurprising that the ALMA sample includes objects that both follow and do not follow the expected size-mass relation for quiescent galaxies.

\begin{figure}[htb]
\includegraphics[width=0.93\columnwidth]{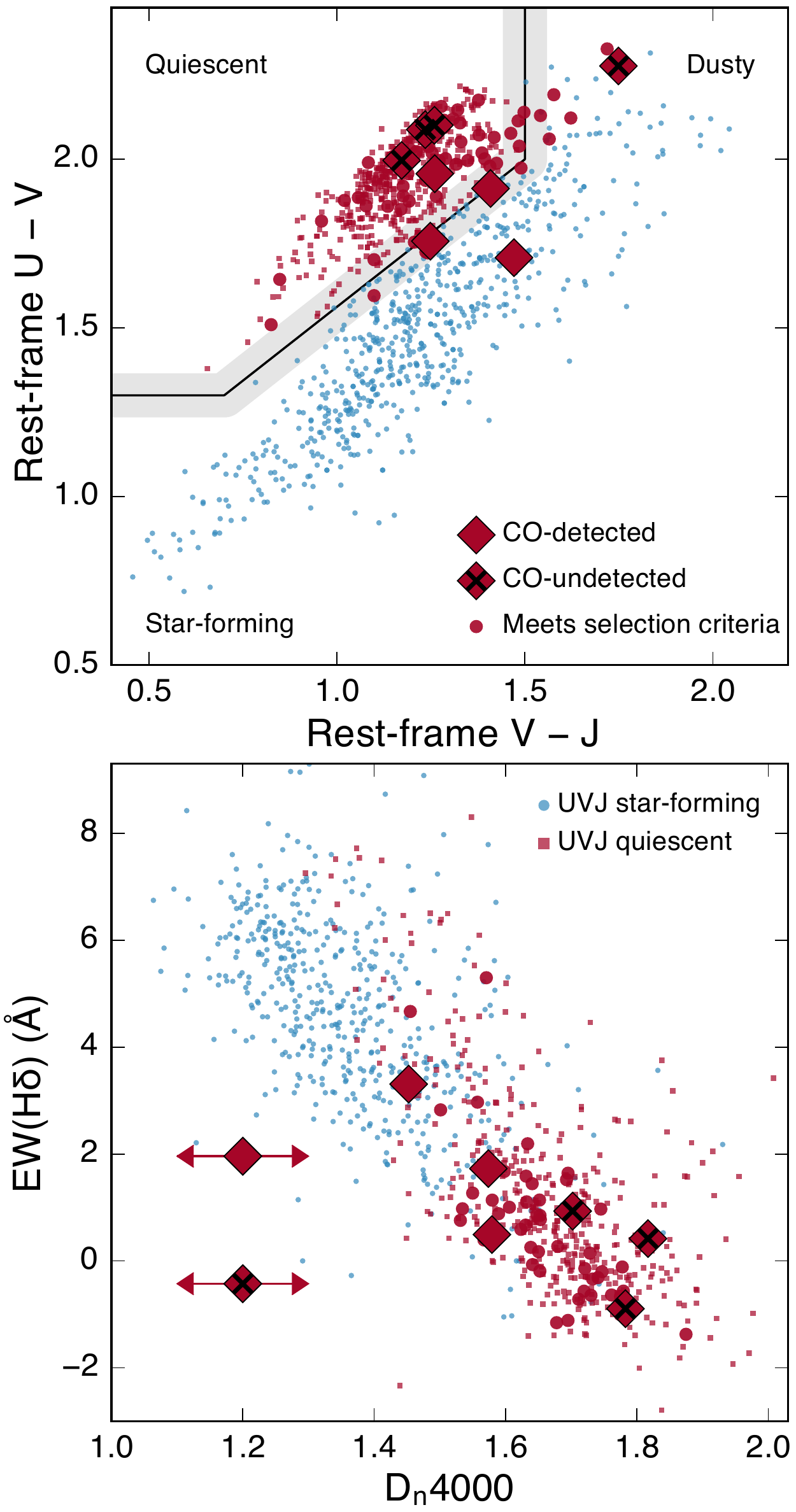}
\caption{
Symbols are plotted as in Figure~\ref{fig:selection}.
\textit{Top:} Rest-frame $UVJ$ color-color diagram for the \legac galaxies and the ALMA sample, with the division between star-forming and quiescent galaxies of \citet{muzzin13a} shown with the black line. The gray shaded band around this line represents both the differences in divisions found in the literature and an expectation that galaxies must transition from one region to the other over a period of time. Although this diagram was not used in the selection of the ALMA sample, six of the eight targets lie within (given the uncertainties) the quiescent region of the diagram at upper left.
\textit{Bottom:} The D$_n$4000 index against the H$\delta$ equivalent width, a proxy for the age of the stellar populations, with older and more massive systems located towards the lower right. We infer typical stellar ages of 1--3\,Gyr for the ALMA sample; these objects are not recently quenched. For two objects (IDs 110509 and 169076), the \legac spectra do not extend sufficiently blueward to measure D$_n$4000, shown arbitrarily at D$_n$4000 $=1.2$ with arrows in each direction.
}
\label{fig:selection2}
\end{figure}

Figure~\ref{fig:selection2} shows two other relevant diagrams that give the ALMA sample additional context. The top panel shows the ALMA sample in $UVJ$ color-color space, with the division between star-forming and quiescent objects from \citet{muzzin13a} also shown. The boundary between these two populations in $UVJ$ colors is not a definitive one, with varying definitions found in the literature as well as an expectation that galaxies must transition from one region to the other over time. Although this diagram was not used in the selection of the ALMA sample, it is reassuring that 6 of the 8 targets lie within (given the uncertainties) the region of the $UVJ$ color-color diagram generally occupied by quiescent galaxies. Of the two remaining objects, one (ID 74512) lies in the region generally occupied by star-forming galaxies; the other (ID 138718) lies in the region generally occupied by dusty galaxies. 

The lower panel of Figure~\ref{fig:selection2} shows the distribution of D$_n$4000 index against the equivalent width of H$\delta$ for the \legac and ALMA-observed samples (\citealt{wu18}; an emission line template has been subtracted from the spectra). This diagram is useful as a proxy for the stellar age, older galaxies showing  lower H$\delta$ equivalent width and higher D$_n$4000. These quantities are also correlated with \Mstar, indicating older stellar populations in massive galaxies \citep[e.g.,][]{kauffmann03,maltby16}. Based on this diagram, we infer typical stellar ages for the ALMA-observed galaxies of 1--3\,Gyr, depending on the assumed star formation history; a similar result is obtained through fitting to the full \legac spectra (Chauke \etal, submitted). In other words, the objects we observed with ALMA have been passive for quite some time. They show little evidence for significant recent star formation, unlike galaxies selected with `post-starburst' criteria that emphasize the recent or impending quenching of star formation \citep[e.g.,][]{tremonti07,sell14,suess17}.

Finally, we examined the local galaxy environments of our target objects using the estimates of \citet{darvish16}. These estimates use photometric redshifts because the spectroscopic completeness in the COSMOS field is low. \citeauthor{darvish16} note that at $z<1$, a comparison of spectroscopic and photometric redshifts shows that the redshift uncertainties are generally small, with a dispersion $\sigma_z\sim0.008$, sufficiently small that the use of photometric redshifts does not wash out line-of-sight galaxy structures. For the \legac sample as a whole, the median overdensity is $\log(1+\delta)=0.13$, while for the ALMA-observed sources, the median is 0.16, with large uncertainties on both values. The ALMA targets are not in systematically overdense environments compared to the full \legac sample or the field as a whole.

%%%%%%%%%%%%%%%%%%%%%%%%%%%%%%%%%% Observations %%%%%%%%%%%%%%%%%%%%%%%%%%%%%%%%%%%%%%
\subsection{ALMA Observations} \label{obs}

Basic details of our ALMA observations and target galaxies are summarized in Table~\ref{tab:targets}. ALMA observations of our sample of 8 galaxies were carried out in project 2016.1.00790.S (PI: Spilker) in separate observing sessions from 17 January to 12 March 2017 using the Band 4 (2\,mm) receivers \citep{asayama14}. All observations were conducted with the array in a compact configuration, with $\sim$40 antennas separated by maximum baselines ranging from 272--330\,m. Observing sessions varied between 57--80 minutes in duration, with 37--49 minutes spent on-source per target. The precipitable water vapor levels varied between 2.7--4.5\,mm, resulting in typical system temperatures of 70--95\,K. Quasars J1058+0133 or J0854+2006 served as bandpass calibrators, while the quasar J0948+0022 was observed for complex gain calibration for all sources. The absolute flux scale was determined using observations of Ganymede or one of J0854+2006 or J1058+0133, both of which are monitored regularly by ALMA. 

The correlator was configured to observe the CO(2--1) line at the known redshift of each target with one baseband with 7.812\,MHz channelization ($\approx 16$\,\kms) after correlator pre-averaging by a factor of 8. Three further basebands with 1.875\,GHz usable bandwidth each were placed to higher frequencies for continuum observations. All data were reduced using the standard ALMA pipeline, with manual inspection of the quality of the reduction.

For each object, we produce continuum images and CO(2--1) spectral cubes with various frequency channelization. All data were imaged with a natural weighting of the visibilities, which maximizes sensitivity to faint emission in exchange for slightly lower spatial resolution. The effective spatial resolution of the data imaged in this way ranges from 1.9--2.4\arc ($\approx$13--17\,kpc). No 2\,mm continuum emission is detected in any target. We extract integrated spectra for all targets by fitting point source models to the visibility data directly, averaging  6 or 12 channels to create spectra with velocity resolution $\sim$100--200\,\kms. When averaging over the full line profiles, we see evidence that most of the detected sources are marginally spatially resolved (e.g., by comparing the peak pixel values with spatially-integrated fluxes), but this effect is negligible in narrower channels.

Finally, because all of our targets are located in the COSMOS extragalactic deep field, we checked the ALMA archive to determine if any other observations of our targets were publicly available or fell within the footprint of other projects. No other observations were found.

\begin{figure*}[htb]
\begin{centering}
\includegraphics[width=0.495\textwidth]{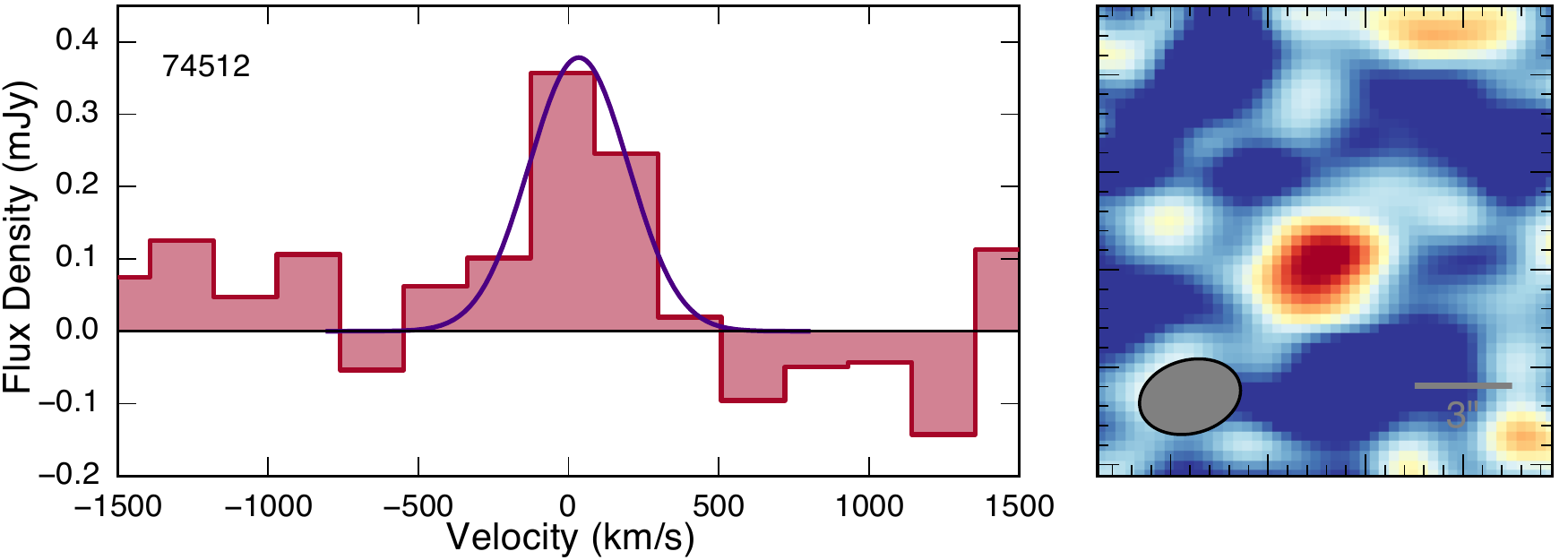}
\includegraphics[width=0.495\textwidth]{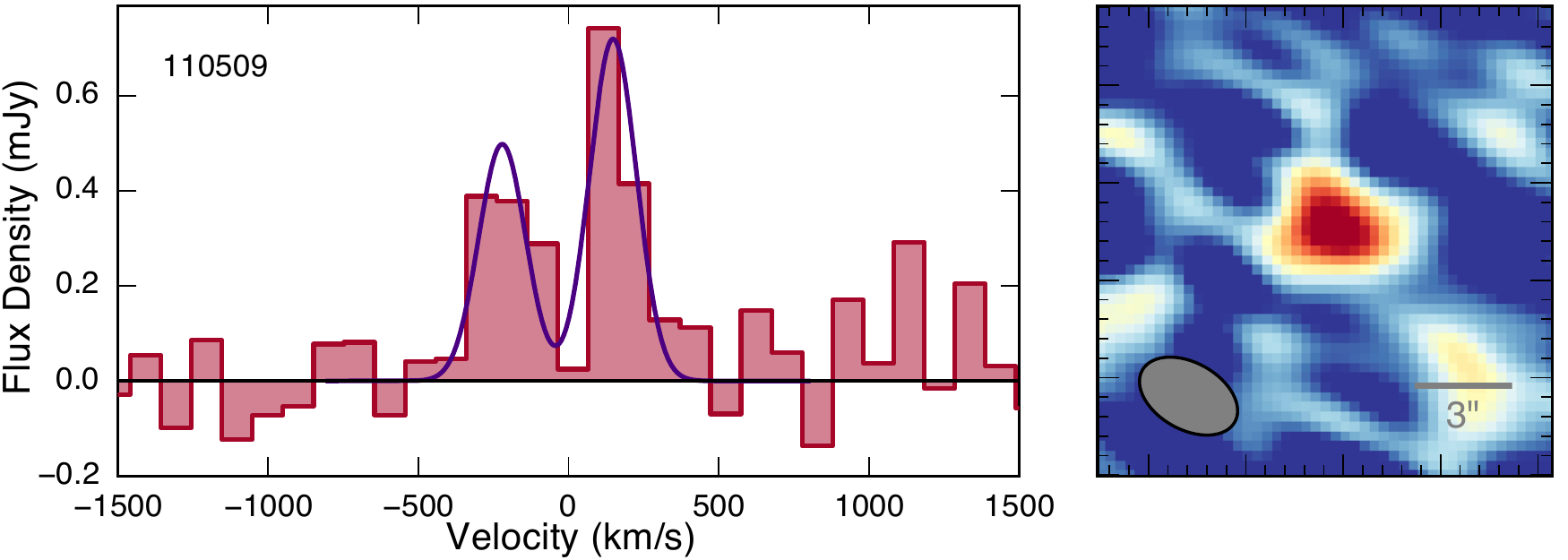}

\includegraphics[width=0.495\textwidth]{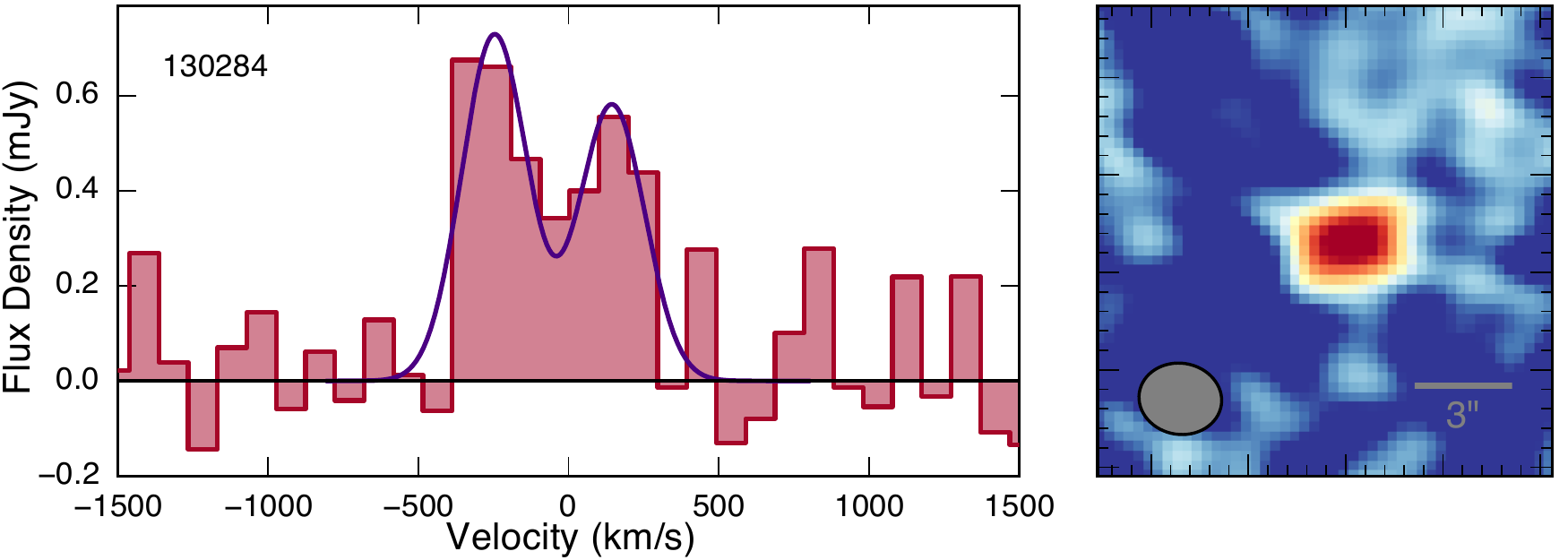}
\includegraphics[width=0.495\textwidth]{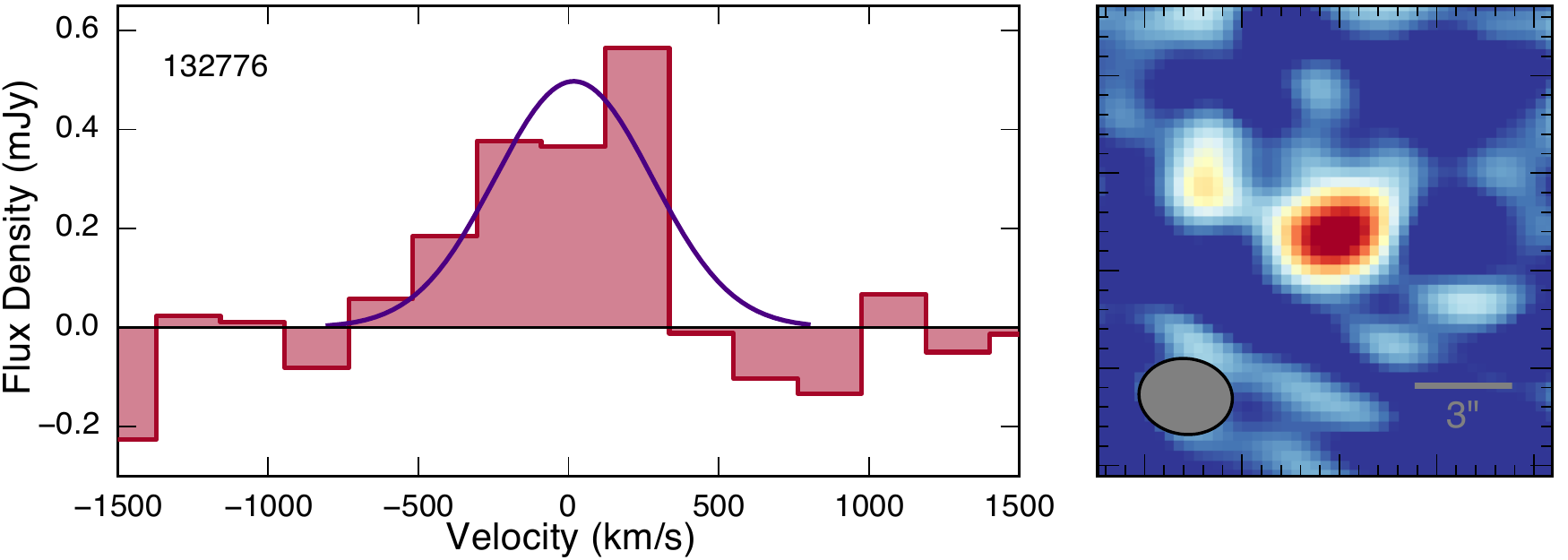}

\includegraphics[width=0.495\textwidth]{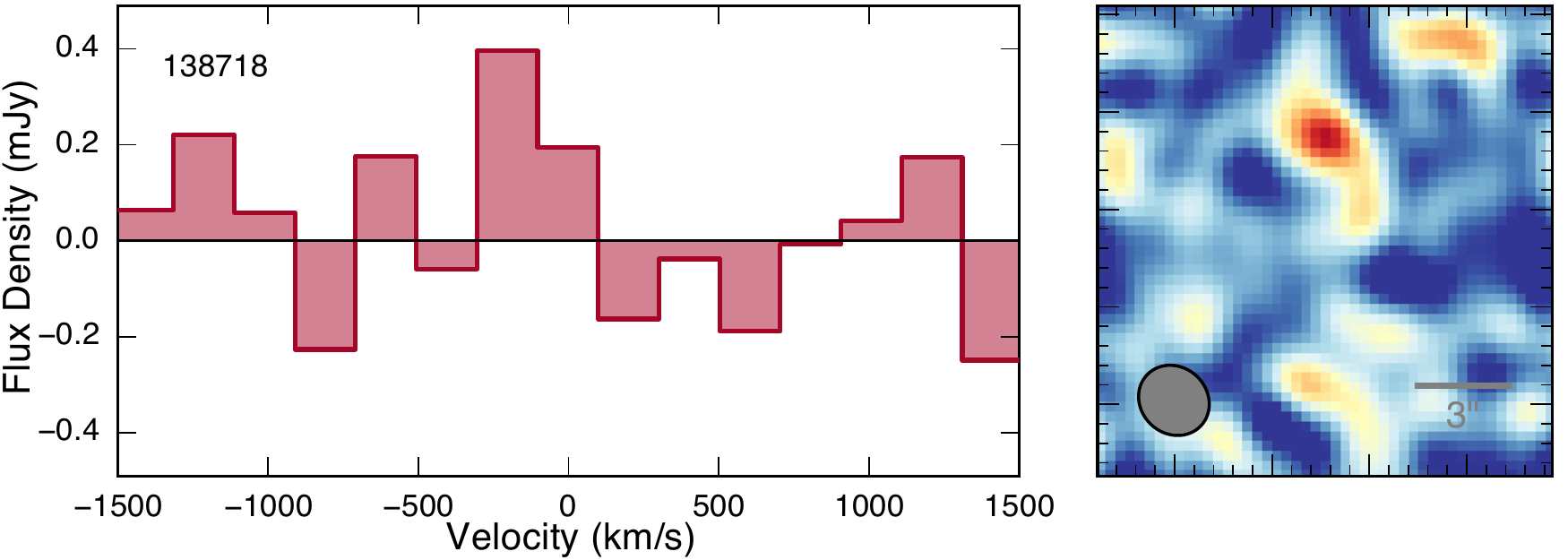}
\includegraphics[width=0.495\textwidth]{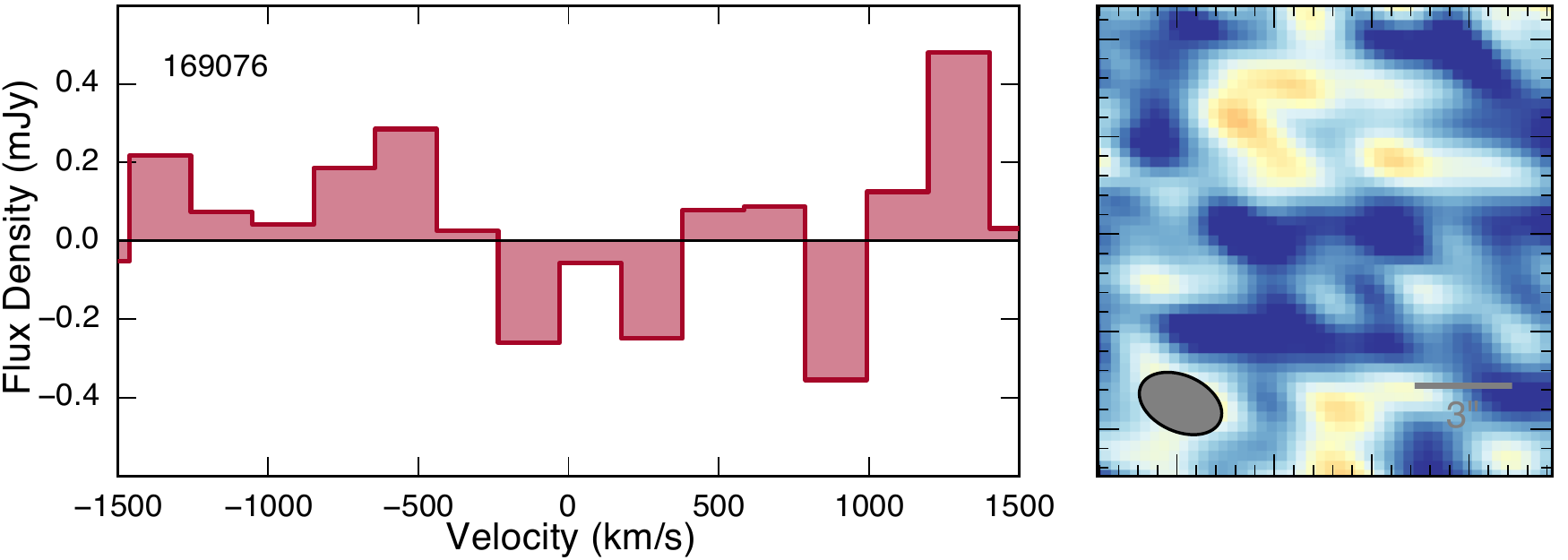}

\includegraphics[width=0.495\textwidth]{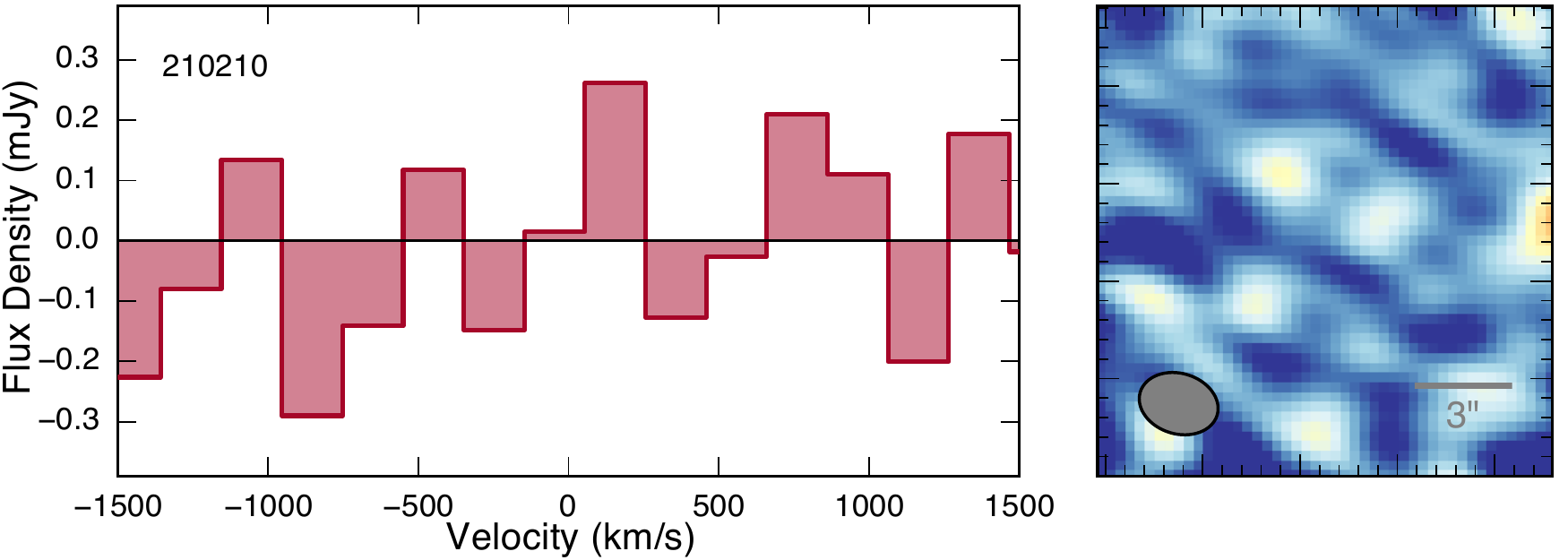}
\includegraphics[width=0.495\textwidth]{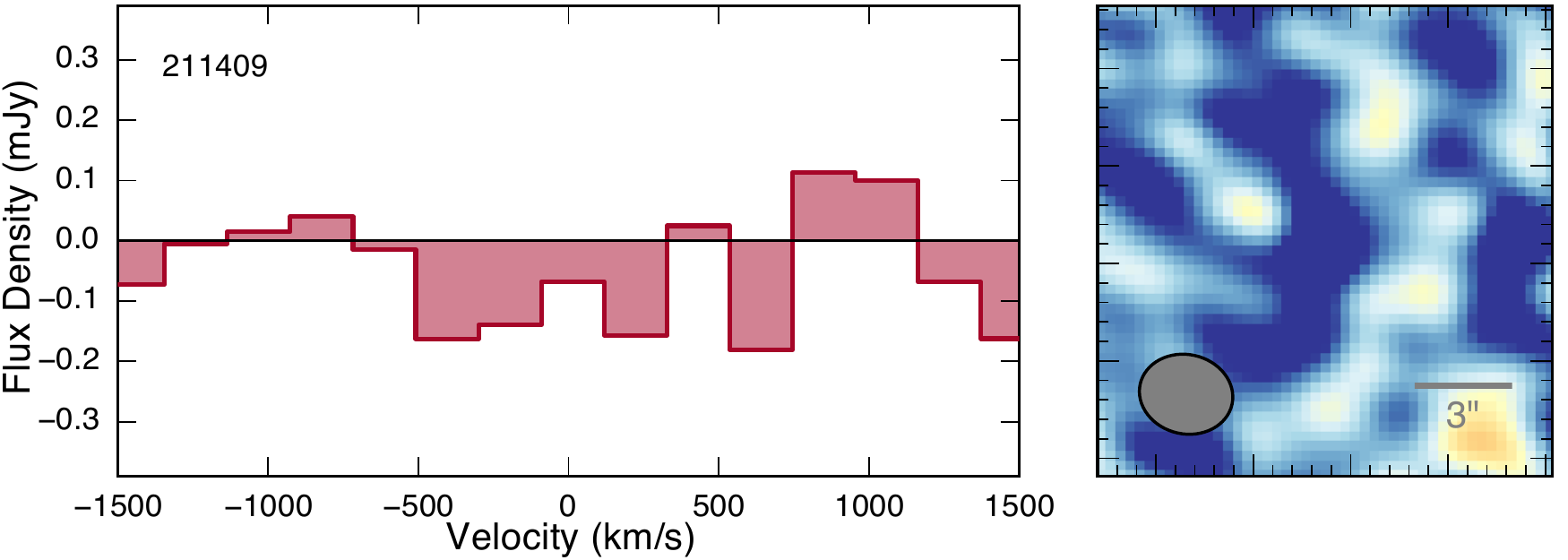}

\end{centering}
\caption{
\textit{Left:} ALMA CO(2--1) spectra for each target. For detected sources, we show
best-fit single or double Gaussian fits with blue lines.
\textit{Right:} Line images integrated over the full line profiles of each source, or 800\,\kms for undetected sources. The grey ovals show the ALMA synthesized beam, and a 3\arc scalebar is indicated.
}
\label{fig:spectra}
\end{figure*}

%%%%%%%%%%%%%%%%%%%%%%%%%%%%%%%%% Mgas, alphaco %%%%%%%%%%%%%%%%%%%%%%%%%%%%%%%%%%%%%%
\subsection{Molecular Gas Masses} \label{gasmethod}

%%%%%%%%%%%%%%%%%%%%% TABLE 1: SUMMARY OF TARGET PROPERTIES ETC %%%%%%%%%%%%%%%%%%%%%
%%%%%%%%%%%%%%%%%%%%%%%%%%%%%%%%%%%%%%%%%%%%%%%%%%%%%%%%%%%%%%%%%%%%%%%%%%%%%%%
% DO NOT EDIT THIS FILE BY HAND, JS HAS A SCRIPT TO GENERATE IT AUTOMATICALLY %
%%%%%%%%%%%%%%%%%%%%%%%%%%%%%%%%%%%%%%%%%%%%%%%%%%%%%%%%%%%%%%%%%%%%%%%%%%%%%%%
\begin{deluxetable*}{rrrrccccc} 
\tablecaption{ALMA-Observed LEGA-C Sample Target Properties\label{tab:targets}}
\tablecolumns{9} 
\tablehead{ 
\colhead{LEGA-C ID} & 
\colhead{Right Ascension} & 
\colhead{Declination} & 
\colhead{z$_\mathrm{spec}$} & 
\colhead{$\log \Mstar / \Msol$} & 
\colhead{SFR} & 
\colhead{$\sigma^{a}_{100\,\kms}$} & 
\colhead{S$_\mathrm{CO(2-1)}$$\Delta v$} & 
\colhead{$\log \Mht / \Msol$} \\ 
\colhead{---} & \colhead{---} & \colhead{---} & \colhead{---} & 
\colhead{---} & \colhead{\Msol/yr} & \colhead{$\mu$Jy/beam} & \colhead{Jy\,\kms} & {---} }
\startdata 
74512   & $10^\mathrm{h}01^\mathrm{m}42.88^\mathrm{s}$   & $+02^\circ01{}^\prime21.9{}^{\prime\prime}$  & 0.7330  & 10.96  & 6.3  & 132   & 0.16 $\pm$ 0.04    & 9.82 $\pm$ 0.13 \\
110509  & $10^\mathrm{h}01^\mathrm{m}04.44^\mathrm{s}$   & $+02^\circ04{}^\prime37.2{}^{\prime\prime}$  & 0.6671  & 11.00  & 6.5  & 160   & 0.24 $\pm$ 0.04    & 9.92 $\pm$ 0.07 \\
130284  & $10^\mathrm{h}00^\mathrm{m}13.78^\mathrm{s}$   & $+02^\circ19{}^\prime37.0{}^{\prime\prime}$  & 0.6017  & 10.96  & 6.8  & 151   & 0.36 $\pm$ 0.04   & 10.00 $\pm$ 0.06 \\
132776  & $10^\mathrm{h}00^\mathrm{m}12.43^\mathrm{s}$   & $+02^\circ21{}^\prime21.9{}^{\prime\prime}$  & 0.7500  & 10.98  & 7.9  & 163   & 0.33 $\pm$ 0.07   & 10.16 $\pm$ 0.11 \\
138718  & $10^\mathrm{h}00^\mathrm{m}13.89^\mathrm{s}$   & $+02^\circ25{}^\prime38.0{}^{\prime\prime}$  & 0.6558  & 11.13  & 5.6  & 188           & $<$0.21            & $<$9.84 \\
169076  & $09^\mathrm{h}59^\mathrm{m}07.30^\mathrm{s}$   & $+02^\circ19{}^\prime05.8{}^{\prime\prime}$  & 0.6772  & 11.49  & 5.1  & 256           & $<$0.23            & $<$9.91 \\
210210  & $10^\mathrm{h}00^\mathrm{m}35.55^\mathrm{s}$   & $+02^\circ31{}^\prime04.2{}^{\prime\prime}$  & 0.6544  & 11.38  & 3.6  & 212           & $<$0.21            & $<$9.84 \\
211409  & $10^\mathrm{h}01^\mathrm{m}05.45^\mathrm{s}$   & $+02^\circ32{}^\prime03.7{}^{\prime\prime}$  & 0.7140  & 11.13  & 6.6  & 188           & $<$0.13            & $<$9.72 \\
\multicolumn{3}{c}{Stacked Non-detections} & 0.6754 & 11.31 & 5.3 & 106 & $<$0.093 & $<$9.51 \\ 
\enddata 
 \tablenotetext{a}{ALMA rms sensitivity in 100\,\kms channels, naturally-weighted images}
\tablecomments{LEGA-C ID numbers are the same as in the UltraVISTA
catalog of \citet{muzzin13a}. Stellar masses are determined by fitting to
multiwavelength photometry using FAST. SFRs are based on a weighted sum of
UV and IR (24\,\um) fluxes. Integrated CO(2--1) line fluxes are converted to
molecular gas masses under the assumptions described in Section~\ref{gasmethod}.
Upper limits for non-detections are 3$\sigma$, and molecular gas masses can be rescaled
under different assumptions as $\Mht (0.8/r_{21})(\alphaco/4.4)$.}
\end{deluxetable*}

Spectra and integrated line images of each target are shown in Figure~\ref{fig:spectra}. We clearly detect four of the eight targets in CO(2--1) emission. Two of the four detections show double-peaked line profiles, usually indicative of rotating disks; the other two sources are centrally-peaked. We discuss kinematics in more detail in Section~\ref{dynamics}. Integrated line fluxes are determined by fitting either one or two Gaussian profiles, also shown in Figure~\ref{fig:spectra}. For the undetected sources, we estimate upper limits on the CO emission by determining the noise in channels 800\,\kms wide, which would fully encompass all of the line emission in all detected sources. The upper limits on the integrated line flux are proportional to $\sqrt{\Delta v}$, where $\Delta v$ is the velocity interval over which the spectrum is integrated, so using very wide channels results in conservative upper limits. The integrated line fluxes are given in Table~\ref{tab:targets}.

We convert the observed CO luminosities and upper limits to estimates of the molecular gas masses under standard assumptions about the CO excitation and the CO--H$_2$ conversion factor \alphaco. The effects of the unknown CO line excitation are minimal, as the CO(2--1) transition we have observed is very close to the ground state CO(1--0) line typically used for molecular gas estimation. Observations of the Milky Way, nearby quiescent and star-forming galaxies, and high-redshift star-forming galaxies ubiquitously show that a line ratio $r_{21}=0.7-1.0$ (in brightness temperature units) encompasses the plausible expected range \citep{fixsen99,combes07,dannerbauer09,young11,spilker14,saintonge17}. In this work, we assume $r_{21}=0.8$.

The subsequent conversion between CO luminosity and molecular gas mass is also uncertain (for a recent review, see \citealt{bolatto13}). The conversion factor \alphaco is known to vary with the gas metallicity \citep[e.g.,][]{leroy11}, which affects the formation and destruction of CO molecules. The metallicities (either gas-phase or stellar) of our targets have not been measured, but are expected to be solar or near-solar based on the mass-metallicity relation. For example, \citet{gallazzi14} predict $\log Z/Z_\odot \sim 0.0-0.1$ for quiescent galaxies in our mass range, with a scatter of $\approx 0.2$. We therefore expect only minor variations in \alphaco due to metallicity effects. The CO-H$_2$ conversion factor also depends on the gas conditions and kinematics, which affect the optical depth of the CO transitions through radiative trapping. This is chiefly relevant for mergers and other high-SFR systems, in which increased gas turbulence and/or bulk motions lower the effective CO optical depth and also \alphaco \citep[e.g.,][]{narayanan12,spilker15}. These effects are also expected to be minor for our targeted objects, which have low SFRs, no signs of interaction, and evidence for disk-like rotation in many cases (see Section~\ref{dynamics}).

In this work, we adopt a `Milky Way-like' value, $\alphaco = 4.4$\,\Msol\,(K\,\kms\,pc$^2$)$^{-1}$ \citep[e.g,][]{solomon87,bolatto13,sandstrom13}. This value agrees with standard dust-based methods we describe further as part of our stacking analysis below. While this choice is justified for the reasons already mentioned, it does still carry significant systematic uncertainty, likely of order 50\%. While this may result in adjustments to the absolute values of the gas masses we derive, the relative values are robust, and it does not affect the overall trends we find. Of relevance to our subsequent discussion in Section~\ref{discussion}, the true gas masses of our sample are unlikely to be significantly larger than the values we infer. The molecular gas masses we derive can easily be rescaled using different assumptions, as $\Mht (0.8/r_{21})(\alphaco/4.4)$. When comparing to other samples and galaxies observed in CO by other authors, we also adjust their derived gas masses to match our adopted value of \alphaco. As we are interested in normal star-forming and quiescent galaxies, this adjustment is minor, no larger than $\sim20$\% for the comparison samples.

%%%%%%%%%%%%%%%%%%%%%%%%%%%%%%%% Stacking Method %%%%%%%%%%%%%%%%%%%%%%%%%%%%%%%%%%%%%
\subsection{ALMA Stacking Analysis} \label{stack}

With CO non-detections constituting half the observed sample, it is worth considering whether these sources are detected on average through a stacking analysis. This would imply that somewhat deeper integrations would have been necessary to detect the objects individually. Similarly, although we did not detect (nor expect to detect) dust continuum emission in any individual target, we expect that it should be detectable in a stacked continuum image, if the assumptions made about \alphaco in the previous section are correct.

\subsubsection{2\,mm Continuum Stack}
Because all targets were observed in similar array configurations with similar synthesized beam sizes, for similar durations, and at similar sky frequencies, we opt to simply average together inverted images of the targets rather than perform this analysis in the Fourier plane. The similarity in observed frequency is of particular importance for the continuum stacked image, minimizing the effects of the steep thermal dust spectral index. We create a stacked 2\,mm continuum image simply by averaging together the signal-free continuum maps of each source. In the stacked image, a faint source with $\stwo = 17 \pm 5$\,\uJy is detected at the center of the field. By subdividing the stack into sources either detected or undetected in CO(2--1), it is apparent that the signal seen in the full stack is due entirely to the sources individually detected in CO. The CO-detected sources are detected in the 2\,mm continuum stack with $\stwo = 23 \pm 7$\,\uJy, while no continuum emission is detected in the stack of CO non-detections, with a 3$\sigma$ upper limit $\stwo < 25$\,\uJy.

We can use the detection of 2\,mm continuum emission in the stack of CO-detected sources to cross-check our assumptions about the CO-H$_2$ conversion factor described in Section~\ref{gasmethod}. The average molecular gas mass of the detected objects is $1 \times 10^{10}$\,\Msol. Assuming a typical dust emissivity, $\kappa_{235\mathrm{GHz}} = 0.42$\,cm$^2$\,g$^{-1}$ \citep[e.g.,][]{dunne00}, where 235\,GHz is the rest-frame continuum frequency, a dust temperature $\Tdust = 25$\,K, and a gas-to-dust ratio typical of solar metallicity systems, $\gdr = 100$ \citep[e.g.,][]{sandstrom13}, we find an average molecular gas mass for the CO-detected sources of $8.8 \times 10^9$\,\Msol. This value is in excellent agreement with our CO-based estimates. The same assumptions applied to the 2\,mm upper limit from the CO-undetected sources yields a 3$\sigma$ upper limit of $\Mht < 9.6 \times 10^9$\,\,\Msol, unsurprisingly in agreement with the CO-based upper limits for individual sources. We note that the dust-based estimate of \Mht is also uncertain by about a factor of two due to uncertainties in the dust emissivity and mass-weighted temperature \citep[e.g.,][]{draine07}. This independent check indicates that our assumptions about \alphaco are reasonable.

\subsubsection{CO Spectral Stack}
We create a stacked CO image cube in similar fashion to the simpler stacked continuum maps. In this case, however, we rely on the fact that all targets (detected and undetected) have precisely-known redshifts from the \legac spectra. The redshifts are accurate to $\sim10$\,\kms, much less than the typical line widths seen in CO or the stellar absorption features in the \legac spectra. We also do not see significant velocity offsets between the CO emission and stellar absorption features in the CO-detected sources. We stack the spectra of individual sources using image cubes with velocity resolution ranging from 50--800\,\kms; the choice of channel width does not affect our conclusions. We have also verified that this procedure accurately recovers the sample-average line flux by stacking only the sources individually detected in CO.

No CO(2--1) emission is detected in the stack of individually-undetected sources, regardless of the velocity resolution used in the stack. As with the individually-undetected sources, we place an upper limit on the CO emission in this stack using a single 800\,\kms wide channel. This results in a 3$\sigma$ upper limit on the CO luminosity $\lprime < 5.8 \times 10^8$\,K\,\kms\,pc$^2$. Under the same assumptions about the CO excitation and \alphaco as before, this is equivalent to $\Mht < 3.2 \times 10^9$\,\Msol. Given the average stellar mass of the CO-undetected sources, $2\times10^{11}$\,\Msol, the resulting upper limit on the gas fraction of these passive sources is $\fht=\Mht/\Mstar<0.016$.

%%%%%%%%%%%%%%%%%%%%%%%%%%%%%%%%%%%%%%%%%%%%%%%%%%%%%%%%%%%%%%%%%%%%%%%%%%%%%%%%%%%%%
%%%%%%%%%%%%%%%%%%%%%%%%%%%%%%%%% Basic Results %%%%%%%%%%%%%%%%%%%%%%%%%%%%%%%%%%%%%
%%%%%%%%%%%%%%%%%%%%%%%%%%%%%%%%%%%%%%%%%%%%%%%%%%%%%%%%%%%%%%%%%%%%%%%%%%%%%%%%%%%%%
\section{Results} \label{results}
\subsection{Basic Gas Properties} \label{basicresults}

\begin{figure*}[tbh]
\begin{centering}
\includegraphics[width=\textwidth]{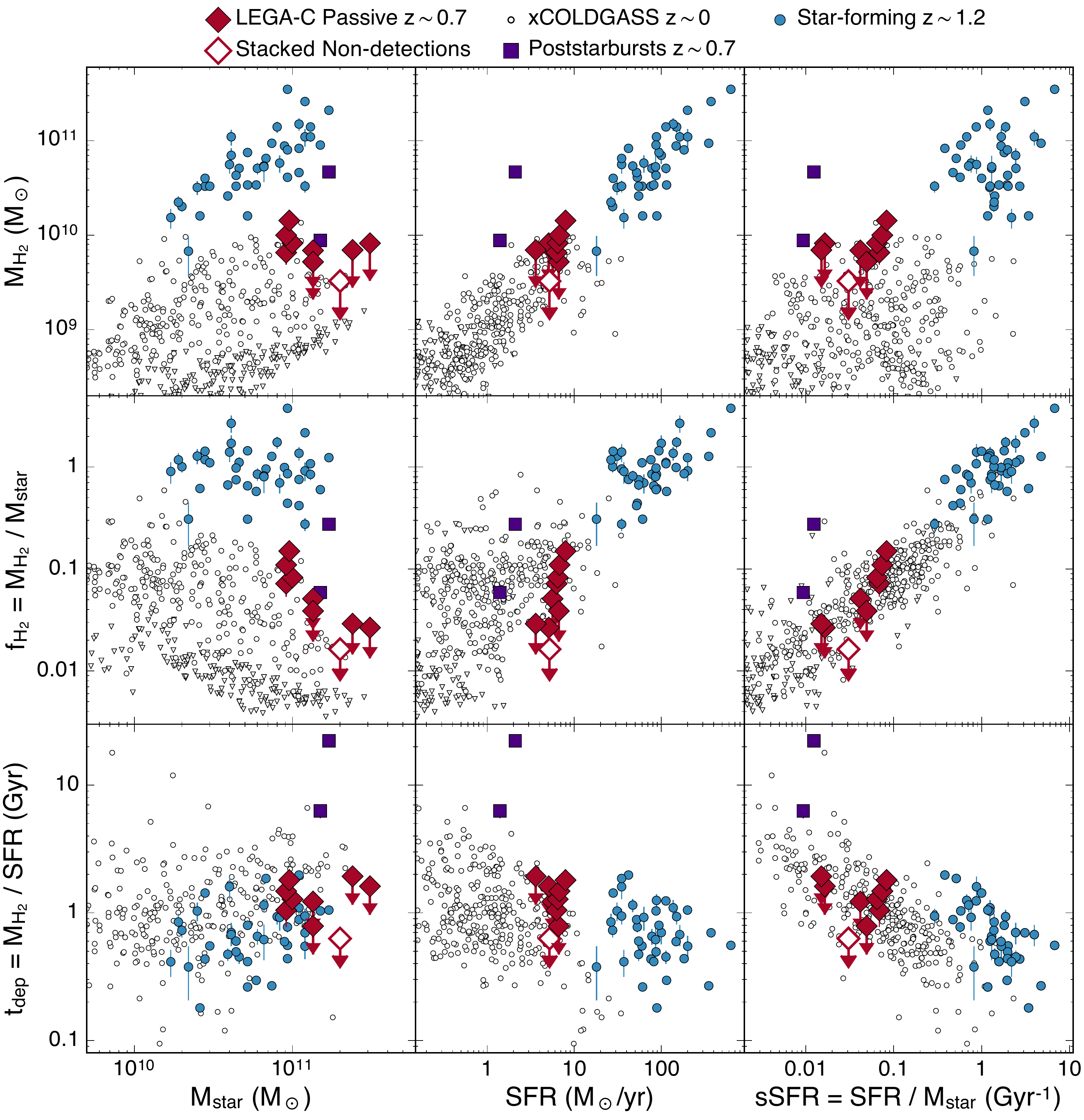}
\end{centering}
\caption{
Summary of basic results derived from our data, with comparison samples detailed in the text. From top to bottom, each row plots the molecular gas mass \Mht, gas fraction \fht, and gas depletion time \tdep; from left to right each column shows the stellar mass \Mstar, star formation rate SFR, and specific SFR. The comparison samples are drawn from \citet{tacconi13,papovich16,saintonge17,suess17}. All upper limits are 3$\sigma$. For clarity of presentation, we do not show upper limits for the xCOLDGASS sample in the bottom (\tdep) row.
}
\label{fig:basicresults}
\end{figure*}

The derived properties of our sample of $z\sim0.7$ passive galaxies are given in Table~\ref{tab:targets}. Figure~\ref{fig:basicresults} plots a number of basic correlations from our data, comparing galaxy molecular masses, gas fractions, and depletion times with their stellar masses, SFRs, and specific SFRs. We draw comparison samples from the literature. We include the xCOLDGASS sample \citep{saintonge17}, a large sample of galaxies in the local universe that contains both star-forming and passive objects over several orders of magnitude in stellar mass. We also include higher-redshift star-forming galaxies at $z=1-1.3$ compiled from the PHIBSS survey \citep{tacconi13}, consisting of $\log \Mstar/\Msol > 10.4$ galaxies, supplemented by the observations of \citet{papovich16}, which extend to lower mass ($\log \Mstar/\Msol \sim 10.2$). Finally, we include the recent ALMA observations by \citet{suess17} of two $z\sim0.7$ galaxies selected to have post-starburst-like optical spectra, $\log \Mstar/\Msol \approx 11$, and little ongoing star formation.  With similar stellar masses to our own sample, these galaxies represent objects that quenched recently, as opposed to the several-Gyr-old stellar populations in our sample. Gas masses from all samples have been renormalized to $\alphaco = 4.4$, but we preserve the original authors' assumptions about the CO excitation, applicable to the $z>0$ samples.

A number of well-known strong correlations are immediately apparent in Figure~\ref{fig:basicresults}, including the spatially-integrated Schmidt-Kennicutt star formation relation between SFR and \Mht \citep{schmidt59,kennicutt98b} and the declining depletion times observed for galaxies with high sSFR \citep[e.g.,][]{saintonge17}. Because all of these quantities evolve with redshift, it is generally more instructive to interpret the variations seen in Figure~\ref{fig:basicresults} after removing or otherwise accounting for the redshift evolution. We perform this exercise in Sections~\ref{scalings} and~\ref{eagle}.

At the most basic level, we observe that $z\sim0.7$ massive, passive galaxies contain $\lesssim10^{10}$\,\Msol of molecular gas, and may contain significantly less, depending on the true masses of the undetected sources. This is nevertheless over an order of magnitude higher \Mht (also \fht) than observed for early-type and other massive, low-sSFR galaxies in the local universe \citep[e.g.,][]{combes07,young11,davis16}. Unlike the local samples, however, we see no evidence for increased gas depletion times that might indicate that the molecular material has been stabilized against collapse; instead, the $\lesssim$1--2\,Gyr depletion times are typical of, or even shorter than, galaxies near the star-forming sequence. 

In terms of basic molecular gas properties, the galaxies we have observed are not so dissimilar from `normal' star-forming galaxies in the local universe. While they were selected to be massive galaxies, the absolute values of \Mht, \fht, and \tdep in the `passive' $z\sim0.7$ galaxies are very similar to star-forming galaxies today. In some sense, the galaxies we have observed can be considered ahead of their time -- they formed the bulk of their stellar mass at early times, and reached an evolutionary stage similar to that currently experienced by galaxies in the local universe $\sim6$\,Gyr early.

It is rather curious that the detected sources are very clearly detected, at the $\sim8\sigma$ level, while the others are not detected even in a stacked spectrum. Why were some objects in our sample detected, and others not? With only 8 target galaxies in total, we cannot draw strong conclusions on this point. The four non-detections are both the highest-\Mstar and lowest sSFR galaxies in the sample, although the absolute SFRs are not significantly different from the full sample. The non-detections tend to lie towards higher stellar surface density than the detections (Figure~\ref{fig:selection}), but are not obviously disparate in morphology or other structural parameters (e.g., S\'ersic index). They also have the lowest H$\delta$ equivalent widths in the sample and the reddest $U - V$ colors (Figure~\ref{fig:selection2}), but do not show large differences in best-fit stellar age, presence or lack of radio AGN, or presence or lack of emission lines in the \legac spectra. Finally, the undetected sources are in regions more overdense than the detected sources by $0.25 \pm 0.4$\,dex.  There is no statistical difference in the relative overdensity between detected and undetected sources. We stress that a much larger sample is required to begin to understand which of these properties, if any, are good or useful predictors of CO line luminosity. 

We see an interesting contrast between our massive, passive galaxies, and the post-starburst galaxies at similar redshift and \Mstar observed by \citet{suess17}. While SFRs for the post-starburst sample are measured from the extinction-susceptible \oii doublet, and thus may be underestimated, the absolute quantity of molecular gas in these two objects still serves as a useful reference. While one post-starburst has a similar molecular mass to the objects we have studied, the other has $>2\times$ higher \Mht than any of our objects, and more than an order of magnitude higher \Mht than the stack of CO-undetected sources in our sample. If the SFR of this object is confirmed to be as low as inferred, it would indicate a large diversity in the gas masses, fractions, and depletion times among massive galaxies at intermediate redshifts.

%%%%%%%%%%%%%%%%%%%%%%%%%%%%%%%%%%%%%%%%%%%%%%%%%%%%%%%%%%%%%%%%%%%%%%%%%%%%%%%%%%%%%
%%%%%%%%%%%%%%%%%%%%%%%%%%%%%%%%%%%% Dynamics %%%%%%%%%%%%%%%%%%%%%%%%%%%%%%%%%%%%%%%
%%%%%%%%%%%%%%%%%%%%%%%%%%%%%%%%%%%%%%%%%%%%%%%%%%%%%%%%%%%%%%%%%%%%%%%%%%%%%%%%%%%%%
\subsection{Molecular and Stellar Dynamics} \label{dynamics}

The dynamical state of the molecular gas in passive galaxies can be a powerful probe of its origin and the processes that led to the cessation of star formation. In $z\sim0$ early-type galaxies, for example, approximately one in four contain detectable amounts of molecular gas, much more common among the fast rotator galaxies than slow rotator population \citep{young11}.  Among these objects, about 40\% show signs that the molecular gas discs are significantly misaligned (more than 30$^\circ$) with respect to the stellar rotation \citep{alatalo13,davis13}. This has been interpreted as evidence that the molecular gas was gained from external sources after the galaxies quenched star formation, for example from the accretion of gas-rich satellites or cold streams from the intergalactic medium. Misaligned rotation axes can also arise during major mergers if gas makes up a significant fraction of the total mass \citep{lagos18}. The dynamics of the molecular gas can thus provide insight into the origin of this material and its relationship to the physical processes that quenched star formation.

From the integrated spectra alone, two of our CO-detected objects show double-horned line profiles characteristic of rotating gas discs, while the other two show centrally-peaked line profiles. This is similar to the results from local early-type galaxies, for which $>30$\% show evidence of rotation from the CO line profiles alone \citep{young11}. The remaining objects may also contain rotating gas discs, depending on the spatial distribution of the molecular gas and the inclination with respect to the line of sight. 

\begin{figure*}[htb]
\begin{centering}
\includegraphics[width=0.24\textwidth]{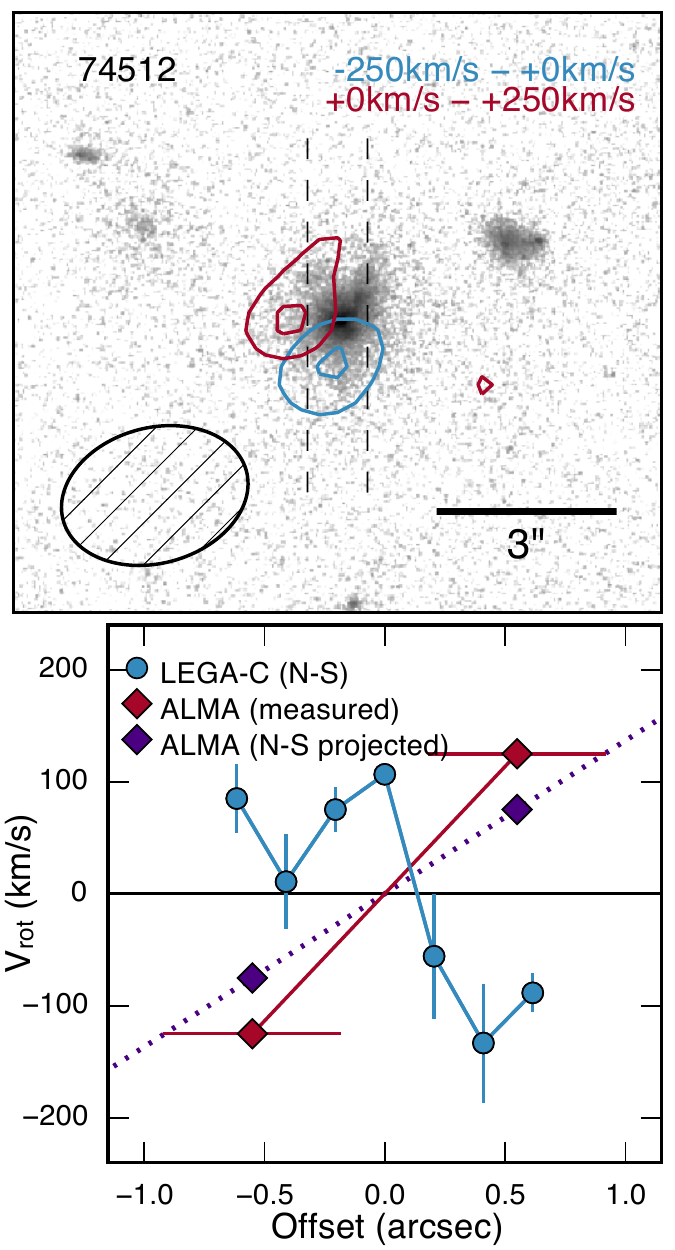}
\includegraphics[width=0.24\textwidth]{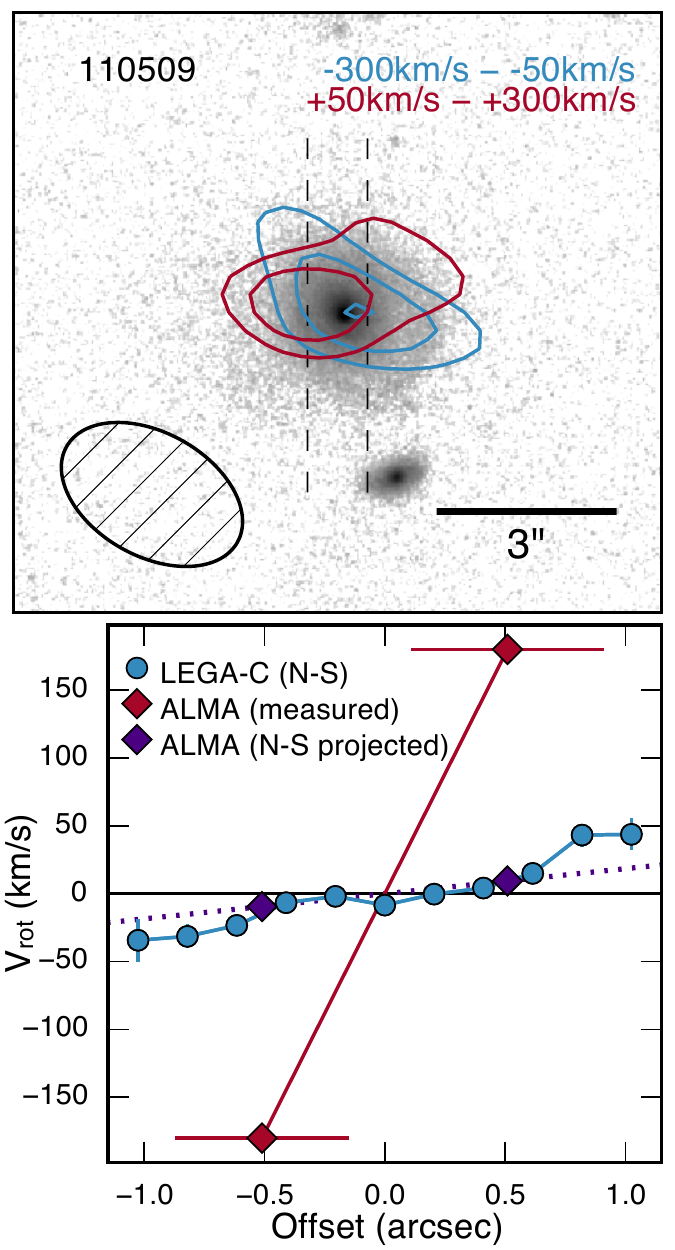}
\includegraphics[width=0.24\textwidth]{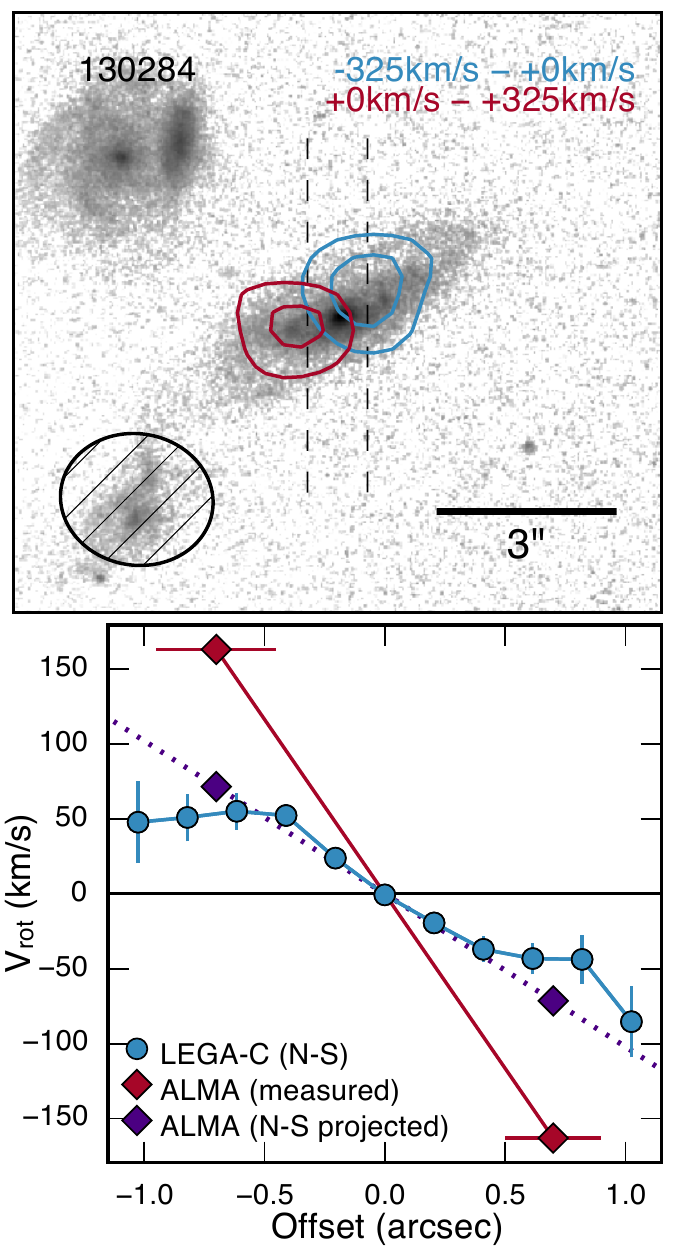}
\includegraphics[width=0.24\textwidth]{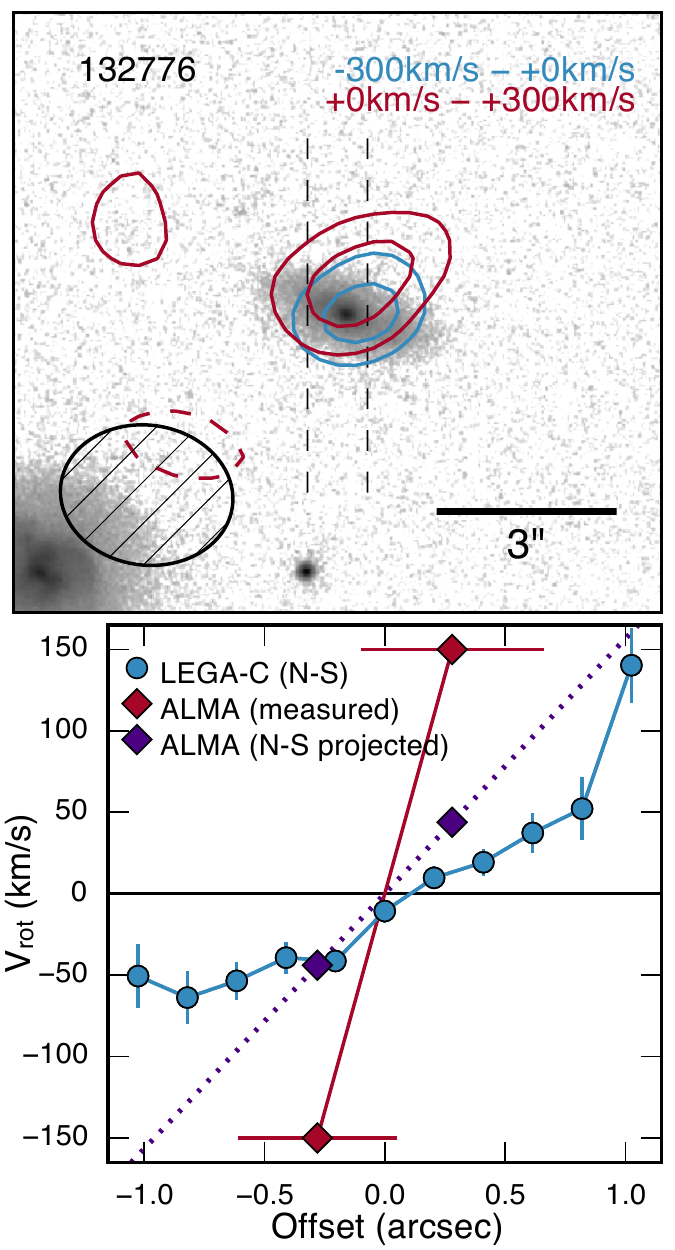}
\end{centering}
\caption{
\textit{Top:} 
Channel maps for each of the targets detected in CO(2--1) emission. The background grayscale shows the \textit{HST}/ACS F814W image of each source, logarithmically scaled. Vertical dashed lines indicate the width and orientation of the \legac VIMOS slits. For each target, we re-image the CO emission in two velocity bins that roughly equally split the total line emission, using the velocity ranges indicated for each source. The blue and red contours show the blue and red velocity components of the CO line in steps of 1$\sigma$ beginning at $\pm$3$\sigma$. The ALMA synthesized beam is shown with an ellipse at lower left; north is up and east is left.  We note that the centroid of each component can be determined to less than a synthesized beam width. Significant velocity gradients are observed in three of four sources; given the modest signal-to-noise, the centroids of the blue and red components of ID 132776 are indistinguishable.
\textit{Bottom:}
CO(2--1) rotation curves derived from the data in the top row (red diamonds). We also show the stellar rotation curves from the \legac spectra (light blue circles), which were all observed with north-south oriented slits. The \legac spectra thus probe only the component of the velocity gradient projected in the north-south direction. The navy diamonds and dotted lines show this component of the rotation curve using the position angle in the ALMA data. In at least three of four cases, the stellar and molecular rotation axes are not obviously misaligned.
}
\label{fig:chanmaps}
\end{figure*}

To investigate the velocity fields of the CO-detected objects in more detail, we re-imaged the ALMA data, splitting the total CO emission from each object into two velocity channels covering the red- and blue-shifted emission. The results are shown in Figure~\ref{fig:chanmaps} (top row). While narrower velocity bins can also be imaged, a single blue and red channel yield the highest signal-to-noise for presentation purposes; we have verified that the velocity gradients apparent in Figure~\ref{fig:chanmaps} are consistent with imaging the data in narrower velocity channels. We fit the red and blue channels of each source in both the visibility and image domains with point source models, finding that the centroids of each velocity component can be determined to $\approx$0.3\arc, on average.  Given the uncertainties, significant velocity gradients are apparent in three of the four detected sources (IDs 74512, 110509, and 130284). The remaining source (ID 132776) may also show a velocity gradient in CO, but the spatial separation between blueshifted and redshifted velocities is not significant at the signal-to-noise of the current data. Our finding that a high fraction of $z\sim0.7$ passive galaxies show measurable rotation mirrors the results for higher-SFR massive galaxies at equivalent and higher redshifts \citep[e.g.,][]{forsterschreiber09,tacconi13,wuyts16}. 

A quantitative comparison between the molecular gas and stellar dynamics is challenging given the data in hand. The \legac spectra provide a high signal-to-noise measurement of the stellar rotation curve \citep{bezanson18}, but all VIMOS slits for the ALMA sample were 1\arc wide and oriented north-south, and so are effectively randomly aligned with respect to the galaxy major axes. Additionally, the typical $\sim$1\arc seeing during the VLT observations leads the rotation curve velocities to be correlated on scales of a kpc. On the other hand, the ALMA CO observations have both lower spatial resolution ($\sim$2--2.5\arc) and lower signal-to-noise compared to the \legac spectra, but contain full two-dimensional spatial information and  independent velocity channelization. 

Given the differences between the available data and the modest signal-to-noise of the current CO maps, we choose not to make a detailed comparison between the CO and stellar dynamics. Instead, we address a simpler question: are the \legac stellar rotation curves consistent with the observed CO velocity gradients, given the misalignment between the projected galaxy rotation axes and the north-south oriented VIMOS slits? In Figure~\ref{fig:chanmaps} (bottom row), we show the velocity gradients seen in the CO data, where the offset between red and blue halves of the emission is measured along the galaxy rotation axis derived from the ALMA data. We also show the \legac stellar rotation curves, which, as mentioned, are sensitive only to the north-south projected component of the velocity field. Finally, we show the portion of the ALMA velocity gradient also projected onto the north-south orientation of the \legac slits.

In three of four cases, we find excellent agreement between the \legac rotation curve and the north-south projection of the CO velocity gradient. The final object, ID 74512, is the most compact galaxy in the sample and does not obviously show signs of rotation in the \legac spectrum. This may be at least in part due to the compact size of the galaxy in comparison to the VIMOS slit and the low signal-to-noise of the stellar rotation curve. It is also possible that the stellar component of this galaxy truly has little rotation, which would imply that the angular momentum axes of the stars and molecular gas are misaligned. Given the large uncertainties involved, however, this scenario is neither supported nor unsupported by the data in hand. 

In summary, the rotational axes of the stellar and molecular components of the CO-detected objects are consistent in at least three of four cases. This argues against an external origin for the molecular gas in $z\sim0.7$ passive galaxies (either recently-accreted cold streams or gas-rich mergers), because these processes should commonly result in misaligned stellar and molecular rotation. Instead, the molecular gas in our sample is probably left over from the formation epoch of the bulk of the galaxies' stellar mass or replenished directly by stellar mass loss, resulting in stellar and gas discs with aligned rotational axes.

%%%%%%%%%%%%%%%%%%%%%%%%%%%%%%%%%%%%%%%%%%%%%%%%%%%%%%%%%%%%%%%%%%%%%%%%%%%%%%%%%%%%%
%%%%%%%%%%%%%%%%%%%%%%%%%%%%%%%%%%%% Scalings %%%%%%%%%%%%%%%%%%%%%%%%%%%%%%%%%%%%%%%
%%%%%%%%%%%%%%%%%%%%%%%%%%%%%%%%%%%%%%%%%%%%%%%%%%%%%%%%%%%%%%%%%%%%%%%%%%%%%%%%%%%%%
\section{Discussion} \label{discussion}
\subsection{Gas Scaling Relations} \label{scalings}

The past decade has seen a large investment of single-dish and interferometer time devoted to understanding how the molecular gas properties of galaxies vary with other galaxy properties from $z\sim0$ \citep[e.g.,][]{bothwell14,saintonge17} out to $z\sim2.5$ \citep[e.g.,][]{tacconi13,genzel15,scoville17,tacconi18}. Because of the well-known correlation between \Mht and SFR (Figure~\ref{fig:basicresults}), the moderate-- to high-redshift samples have focused nearly exclusively on fairly massive star-forming galaxies on or above the star-forming sequence, the objects with the largest gas masses and most easily detectable. With our sample of passive galaxies at intermediate redshifts, we are in a position to determine whether these scaling relations extend to lower sSFR, previously unexplored parameter space.

The above studies provide prescriptions for the variation of molecular gas fraction \fht and depletion time \tdep, parameterized in terms of overall redshift evolution, sSFR (generally with respect to the sSFR expected from the star-forming sequence at a given epoch), \Mstar, and/or galaxy size \reff. Based on samples of hundreds of galaxies from $z=0-3$, consensus has emerged that the gas fraction \fgas evolves steeply with redshift, $\propto$$(1+z)^{1.8-2.5}$, with shallower dependencies on star-forming sequence offset, $\propto$$(\Delta \mathrm{sSFR})^{0.3-0.5}$, and stellar mass, $\propto$$(\Mstar)^{-(0.3-0.7)}$. The evolution of \fgas with redshift appears to be slightly less rapid than the normalization of the star-forming sequence itself. This implies that to first order, the higher SFRs observed in typical galaxies at high redshift are simply due to larger gas masses, with perhaps a somewhat higher efficiency of star formation (equivalently, lower \tdep) also needed. The depletion time \tdep exhibits an overall smaller dynamic range, and varies less steeply with redshift, $\propto$$(1+z)^{-(0.3-1)}$, star-forming sequence offset, $\propto$$(\Delta \mathrm{sSFR})^{-(0.4-0.7)}$, and very shallow dependence on stellar mass, $\propto$$(\Mstar)^{0-0.17}$. Because $\tdep\approx1$\,Gyr for star-forming galaxies, with shallow redshift evolution, galaxies must also have had high gas accretion rates in order to reconcile \tdep with the evolution of \fht, reaching $>100$\,\Msol/yr at $z>2.5$ \citep{scoville17}.  

It is not clear whether these scaling relations should or do extend to passive galaxies significantly below the star-forming sequence. On one hand, the scaling relations are very successful over a very wide parameter space, with residual scatter in \fht and \tdep of just $\sim0.1$\,dex \citep{tacconi18}. On the other hand, the scaling relations have been derived using only star-forming galaxies, and thus may not account for the physical mechanisms that quench galaxies, or the diversity of these mechanisms that may induce increased scatter in the relations.

\begin{figure*}[htb]
\begin{centering}
\includegraphics[width=\textwidth]{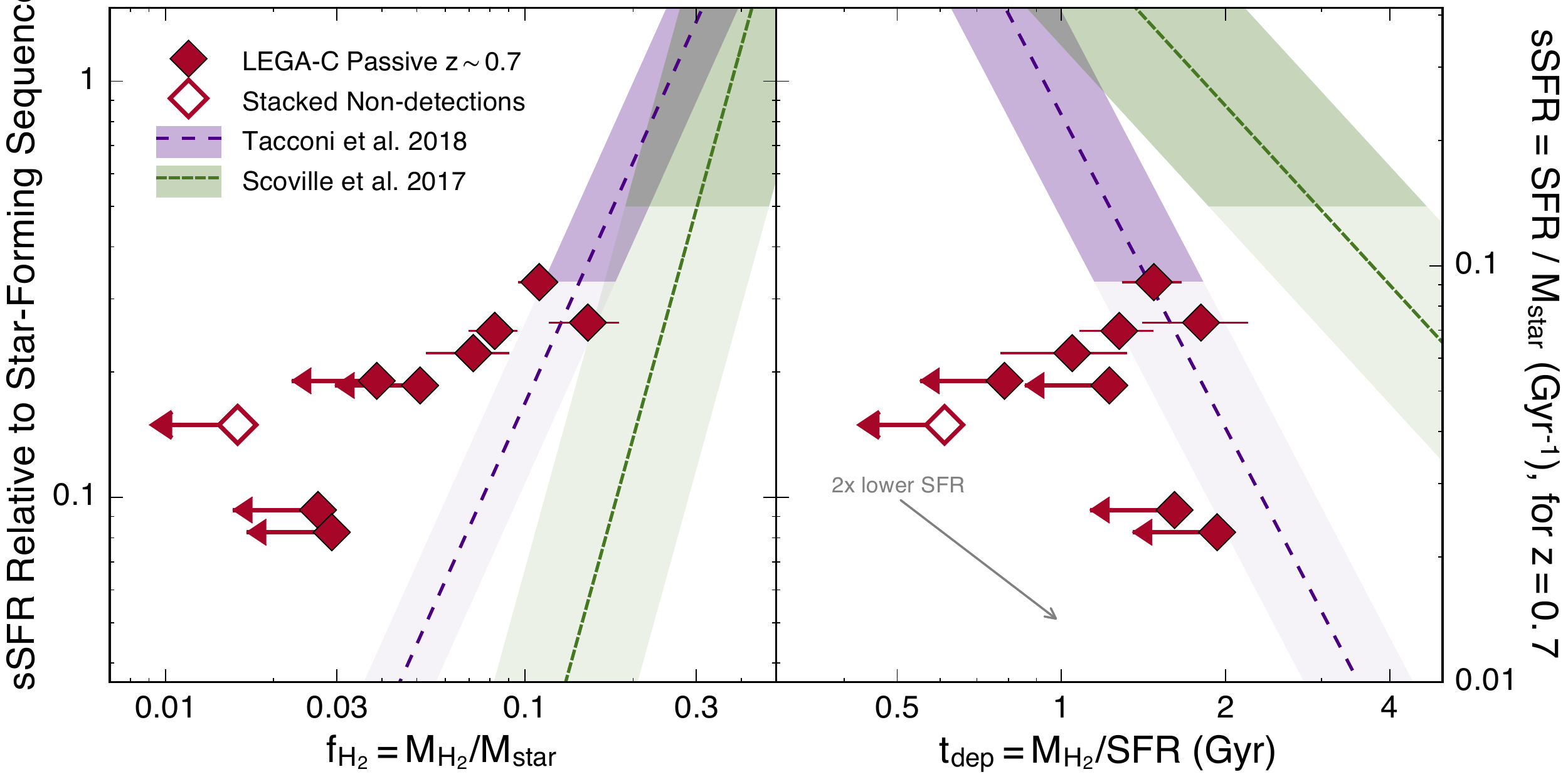}
\end{centering}
\caption{
The molecular gas fraction \fht (\textit{left}) and depletion time \tdep (\textit{right}) versus the specific SFR, compared with observationally-derived literature scaling relations from \citet{scoville17} (green densely dashed line and region) and \citet{tacconi18} (navy dashed line and region), evaluated for $\log \Mstar/\Msol=11$. Both literature scaling relations are parameterized in terms of the offset of the sSFR with respect to the expectation from the star-forming sequence, and so the left-hand y-axis is also labeled in these terms. The right-hand y-axis is labeled with the corresponding values of the sSFR, evaluated at $z=0.7$. The shaded regions encompass the quoted approximate 1$\sigma$ residual scatter around the relations. The darker shaded bands correspond to the approximate sSFR lower bounds for galaxies at $z=0.6-0.8$ that were used in the derivation of the scaling relations; below these regions the scaling relations are extrapolations (lightly shaded). In the right-hand panel, a grey arrow indicates the shift if the SFRs for our galaxies have been overestimated by a factor of 2; the shift is largely parallel to the scaling relations. In general, the CO-detected objects in our sample are in reasonable agreement with the scaling relations, while the undetected objects, and especially the stacked non-detections, are significantly offset from the extrapolated relations. This could indicate either a strong break in the scaling relations at 3--5$\times$ lower sSFR than the star-forming sequence, or a large increase in the scatter of \fht and \tdep.
}
\label{fig:scalings}
\end{figure*}

In Figure~\ref{fig:scalings}, we compare the gas fractions and depletion times for our observed sample with two empirical scaling relations from the literature.\footnote{Both \citet{scoville17} and \citet{tacconi18} provide scaling relations using the prescription for the star-forming sequence of \citet{speagle14}, which we also adopt for Figure~\ref{fig:scalings}. The difference between the \citeauthor{speagle14} and \citet{whitaker12} formulations explains why two objects in our sample are $>$10$\times$ below the star-forming sequence, lower than our nominal selection threshold.} The relations derived by \citet{scoville17} are based on ALMA observations of long-wavelength dust continuum emission and assumptions about the dust emissivity. The work by \citet{tacconi18}, on the other hand, is a meta-analysis of studies that represents the largest overall sample size and dynamic range to date, and also accounts for different normalizations in the major techniques used to measure gas masses. \citet{tacconi18} adopt a metallicity-dependent CO-H$_2$ conversion factor based on the observed mass-metallicity relation, while we simply use $\alphaco=4.4$. We note that using their prescription  for our sample gives a smaller value for \alphaco, resulting in even lower gas masses, fractions, and shorter depletion times; in other words, the differences seen in Figure~\ref{fig:scalings} would be even further accentuated.

The darkly shaded regions in Figure~\ref{fig:scalings} show the approximate lowest sSFR objects at $z\sim0.7$ contained in the observational samples; at lower sSFR (lightly shaded), the scaling relations are extrapolated. For both \fht and \tdep, Figure~\ref{fig:scalings} shows that the extrapolation of the \citet{tacconi18} scaling relation (which also incorporates the data from \citealt{scoville17}) is a better match to our data, while the \citet{scoville17} relations predict significantly higher \fgas and longer \tdep compared to our observations. The \citet{tacconi18} scaling relations perform reasonably well for the CO-detected sources closest to the star-forming sequence. However, both scaling relations perform poorly for the CO-undetected sources in our sample, especially so for the stacked non-detections. Individual non-detections have upper limits on \fgas lower than the extrapolations by $\sim2\times$, while the upper limit on \fgas in the stack is a factor of 5 lower than expected. A similar discrepancy applies to \tdep, with individual sources showing depletion times shorter by $1.6-2.5\times$ and the stack too short by a factor of 4. 

Considering either \fht or \tdep, the CO-undetected sources contain significantly less molecular gas than expected. This result holds even if we have overestimated the SFRs in our sample by $\gtrsim5\times$, because the scaling of \fht with sSFR is very steep, and altering the SFR moves galaxies largely parallel to the \tdep scaling relation. At the sSFRs of our sample, we expect the SFRs to be overestimated by at most about a factor of two (\citealt{fumagalli14}, and Section~\ref{selection}). It is interesting to note that the four undetected sources are the four furthest below the star-forming sequence, but with a small sample size it is unclear that this is significant (Section~\ref{basicresults}).

It is apparent that the literature scaling relations cannot be extrapolated and lose predictive power at the low sSFRs probed by our sample. Instead, there appears to be either a break in the scaling relations towards lower \Mht at sSFRs 4--5$\times$ below the star-forming sequence, or a sharp increase in the scatter of the scaling relations, with our small sample size explaining the lack of detections closer to the expected relations. Equivalently, a curvature term in the $\Delta(sSFR)$ dependency could be required in the scaling relations in order to bring them into agreement with our observations. Only a larger sample of galaxies well below the star-forming sequence would reveal whether this is necessary.

It is not clear why half of the observed sample deviates so significantly from the extrapolated scaling relations. If this trend is real, the answer likely involves the physics of galaxy quenching, including feedback processes and galactic winds, gas starvation or lowered accretion rates, environmental effects, and/or morphological transformation. One or more of these processes likely acted on the observed galaxies in order to disrupt their equilibrium growth, leaving them with significantly less molecular gas than otherwise expected. We return to this question in Section~\ref{shorttdep}.

%%%%%%%%%%%%%%%%%%%%%%%%%%%%%%%%%%%%%%%%%%%%%%%%%%%%%%%%%%%%%%%%%%%%%%%%%%%%%%%%%%%%%
%%%%%%%%%%%%%%%%%%%%%%%%%%%%%%%%%%%%%% EAGLE %%%%%%%%%%%%%%%%%%%%%%%%%%%%%%%%%%%%%%%%
%%%%%%%%%%%%%%%%%%%%%%%%%%%%%%%%%%%%%%%%%%%%%%%%%%%%%%%%%%%%%%%%%%%%%%%%%%%%%%%%%%%%%
\subsection{Comparison to Cosmological Simulations} \label{eagle}

Recent large cosmological simulations have been able to produce populations of galaxies with properties broadly consistent with many observations \citep[e.g.,][]{vogelsberger14,schaye15}. Typical target benchmarks include matching the galaxy stellar mass function at $z=0$ and higher redshifts, the star formation history of the universe, galaxy colors, and other observables. The generally good agreement between the simulations and these basic observed properties then allows us to explore more detailed aspects of the simulations, which can help probe whether the sub-grid physical processes modeled by the simulations are accurately calibrated. Passive galaxies in particular are useful in this comparison, as they are the end products of powerful galactic feedback and quenching processes that encapsulate the most significant differences between various simulations. Here we compare the molecular gas properties of our observed $z\sim0.7$ passive galaxies to predictions from the  Evolution and Assembly of GaLaxies and their Environments (EAGLE) simulation \citep{schaye15}.

Molecular gas masses for galaxies in the Ref-L100N1504 EAGLE run were calculated by \citet{lagos15,lagos16} using two different prescriptions for the molecular fraction that depend on, e.g., the metallicity, density, pressure, and SFR of each gas particle \citep{gnedin11,krumholz13}. In comparing the simulated galaxies' SFRs and gas masses with observations at $z=0$, \citet{lagos15} find good agreement for `normal' star-forming galaxies. On the other hand, EAGLE underpredicts the number of gas-rich early-type galaxies compared to observations; early-type galaxies from the ATLAS$^{\mathrm{3D}}$ survey are over-represented at the gas-rich end compared to the simulation, with 17\% of observed galaxies lying $>$2$\sigma$ from the median SFR--\Mht relation in EAGLE. Although no morphological selection of early types was performed in the simulation, clouding the comparison, the difference could indicate that the star formation `law' employed by EAGLE is too simplistic.

\begin{figure*}[htb]
\centering
\includegraphics[width=\textwidth]{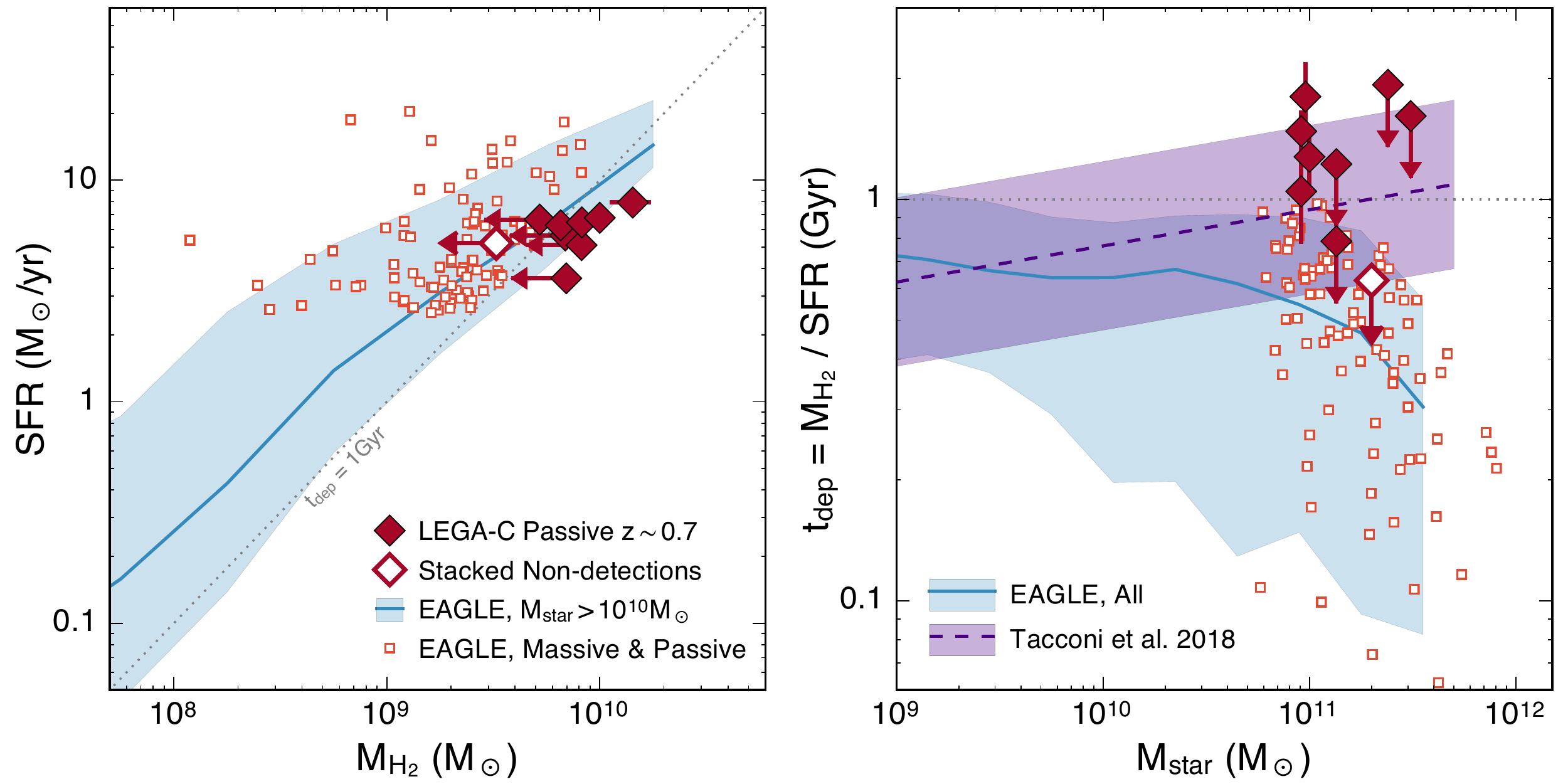}
\caption{
\textit{Left:}
The star formation relation from a $z=0.6$ snapshot of the EAGLE simulation, compared with our results. The individual objects and stacked non-detection from our observations are shown with red diamonds. The blue line and shaded region show the median and 90\% range of all galaxies in EAGLE with $\log \Mstar/\Msol > 10$. The small orange squares show galaxies in EAGLE that mimic our selection criteria; see text for details. The dotted grey line shows a constant depletion time $\tdep=1$\,Gyr. 
\textit{Right:}
The gas depletion times as a function of stellar mass from the same EAGLE snapshot. The blue line and shaded region again shows the median and 90\% range of depletion times for all EAGLE galaxies, with those that mimic our selection criteria as orange squares. The dashed navy line and region, on the other hand, show the predicted scaling relation from \citet{tacconi18} for galaxies on and a factor of 3 above (lower edge of region) and below (upper edge) the star-forming sequence at these redshifts. Note also that this Figure does not contradict our results in Figure~\ref{fig:scalings}, even though the \tdep upper limits from the CO non-detections overlap with the shaded region, because these galaxies have significantly lower sSFR than merely 3$\times$ below the star-forming sequence.
}
\label{fig:eagle}
\end{figure*}

In Figure~\ref{fig:eagle} we make a similar comparison to the simulations using our ALMA observations. This figure shows the SFR--\Mht star formation relation using a $z=0.6$ snapshot from EAGLE; we obtain a virtually identical result using a $z=0.85$ snapshot. For this comparison, we simply average the results from the two prescriptions used by \citet{lagos15} to calculate \Mht for the simulated galaxies; using one or the other alone also has no impact on our conclusions. The blue line and shaded region in Figure~\ref{fig:eagle} show the median and 90\% interval of SFR as a function of \Mht for all galaxies in EAGLE with $\log \Mstar/\Msol > 10$, approximately the stellar mass cutoff of the \legac survey.  

Making a clean selection of simulated galaxies similar to our own selection criteria for ALMA observations is difficult. While we primarily selected objects from the \legac survey to have sSFRs 3--10$\times$ below the star-forming sequence at $z\sim0.7$, it is known that EAGLE does not accurately track the evolution of the star-forming sequence to high redshifts. In particular, the simulated galaxies underpredict the normalization of the star-forming sequence by $\sim$0.2--0.4\,dex at $z=1$ \citep{furlong15}. This issue is not unique to EAGLE -- the same is seen in other cosmological simulations (e.g., Illustris; \citealt{sparre15}). It is also worth noting that even the observations are not self-consistent on this point, with the integral of the cosmic SFR density over-predicting the observed stellar mass density by $\sim0.2$\,dex (e.g., \citealt{madau14}, though see also \citealt{driver18}), while the simulations obviously must be self-consistent. Given the uncertainty in the normalizations of these relations in the literature, instead of selecting galaxies with the same range of sSFRs as the observations, we instead simply select the objects with sSFRs lower than 95\% of the total population of simulated galaxies. Although the quantitative selection is different, this preserves the spirit of our selection, aiming for objects much more passive than the overall population. We also restrict the comparison to galaxies with $\Mht>10^7$\,\Msol, below which gas masses are unreliable due to the baryonic particle mass resolution of EAGLE. We preserve our other two main selection criteria, $\log \Mstar/\Msol > 10.8$ and SFR$>$2.5\,\Msol/yr, as in the observed sample, resulting in 91 simulated massive, passive galaxies. While these are absolute, rather than relative, selection cuts, they have little influence on our conclusions because, as we will show, massive and passive galaxies in EAGLE do not show a systematically different star formation relation compared to the rest of the simulated galaxies. The objects mimicking our selection are shown with small orange squares in Figure~\ref{fig:eagle}.

We find that our observed sample tends to lie near, but slightly offset from the bulk of simulated galaxies. The CO-detected sources, in particular, seem to have either too much molecular gas given their SFRs, or too little star formation given the amount of gas they contain. In other words, their depletion times are slightly longer than seen in the simulation. We draw the same conclusion from comparing only to those objects that mimic our selection criteria, which are not systematically offset from the rest of the galaxies in the simulation. The differences would be further exaggerated if the SFRs in the simulation or observations have been poorly estimated -- either too low in EAGLE due to the underprediction of the star-forming sequence, or too high in our targets due to non-SF dust heating at 24\,\um. If the SFRs of the observed and comparison simulated galaxies are generally correct, however, it appears that the differences are mostly in \Mht, and not in SFR. The simulated galaxies have SFRs comparable to the observed sample (partly by construction, due to our SFR selection threshold), but a median $\Mht = 2\times10^9$\,\Msol, with a dispersion of 0.2\,dex and a tail that extends to lower molecular masses. Indeed, even the most gas-rich EAGLE galaxy matching our selection criteria has lower \Mht than any of our CO-detected objects.

Amusingly, unlike the observationally-derived scaling relations of Section~\ref{scalings}, in this comparison it is the CO-\textit{undetected} \legac objects that are in better agreement with the scaling relation. While we have only upper limits on the depletion time of these objects, they would be consistent with the EAGLE galaxies even if they contained more than an order of magnitude less molecular gas than our present upper limits. 

The better agreement for the non-detections appears to be due to the fact that EAGLE predicts a weak decline of \tdep towards high masses, in contrast to the increasing depletion times for massive galaxies predicted by the \citet{tacconi18} scaling relations. In EAGLE, the decline in \tdep is largely due to two linked effects. First, the probability distribution function of ISM gas densities is shifted to higher mean density in more massive galaxies. Second, EAGLE assumes a super-linear Schmidt-Kennicutt star formation relation. Combined with the higher gas densities in more massive galaxies, this leads to more star formation per unit molecular gas (and lower depletion times) at high masses. This effect would be muted if instead a linear relationship between SFR and \Mht were used in the simulations. EAGLE therefore does not and cannot reproduce the diversity of \tdep values that are unveiled by our observations of passive galaxies, suggesting that adopting a universal Schmidt-Kennicutt relation in the simulation is not sufficient to capture the complexity these observations reveal.

In Figure~\ref{fig:eagle} (right),  we examine the depletion times explicitly, comparing the predicted correlation between \tdep and \Mstar from \citet{tacconi18} with that seen in the EAGLE simulation, where the navy shaded region corresponds to the prediction for galaxies a factor of 3 below (upper edge) and above (lower edge) the star-forming sequence at these redshifts. The two scaling relations predict opposite behavior for high-mass galaxies. The differences are especially stark for massive and passive galaxies (again shown with orange squares), which follow the same envelope as star-forming galaxies in EAGLE but are predicted to have depletion times longer by a factor of $>5\times$ by the \citet{tacconi18} scaling relation.

Our observations, in summary, provide some support for both the observationally-- and theoretically-derived scaling relations. While the gas-rich, CO-detected objects in our sample agree reasonably well with the extrapolations from \citet{tacconi18}, our undetected objects have depletion times that are much more consistent with the EAGLE simulation at these redshifts \citep{lagos15,lagos16}. This highlights the importance of building a larger sample of passive galaxies with molecular mass estimates, which would indicate whether one or both of these scaling relations should be revised -- how these scaling relations should be revised is unclear without additional information from a larger sample.

%%%%%%%%%%%%%%%%%%%%%%%%%%%%%%%%%%%%%%%%%%%%%%%%%%%%%%%%%%%%%%%%%%%%%%%%%%%%%%%%%%%%%
%%%%%%%%%%%%%%%%%%%%%%%%%%%%%%%%%%% Quenching %%%%%%%%%%%%%%%%%%%%%%%%%%%%%%%%%%%%%%%
%%%%%%%%%%%%%%%%%%%%%%%%%%%%%%%%%%%%%%%%%%%%%%%%%%%%%%%%%%%%%%%%%%%%%%%%%%%%%%%%%%%%%
\section{Implications for Galaxy Quenching} \label{quenching}

\subsection{Active and Dynamical Quenching} \label{shorttdep}

In contrast to the nearly-linear relationship between \Mstar and SFR observed for lower-mass galaxies from $z=0$ to $z\approx2.5$, high-mass galaxies ($\log \Mstar/\Msol \gtrsim 10.5$) tend to show a sub-linear relationship \citep[e.g.,][]{whitaker14,schreiber15}. This is generally interpreted as an effect of the `mass quenching' of high-mass galaxies that results in a nearly-constant break in the stellar mass function at $\log \Mstar/\Msol \sim 10.5-11$ \citep{peng10b}. The various physical mechanisms that cause this mass quenching predict different observable properties of the ISM in passive galaxies, and can potentially be disentangled by observations of large samples of massive, quenched objects such as those presented here.

At $z\lesssim1$, the local environment of a galaxy is correlated with whether or not it is quenched, with galaxies in dense regions much more likely to be quiescent than those in less dense environments \citep[e.g.,][]{sobral11,darvish16}. Numerous physical mechanisms can explain this environmental dependence, including ram pressure stripping, starvation, and tidal interactions. The effects of `maintenance-mode' AGN feedback are also most prominent in denser environments \citep{correa18}. We found no difference in local overdensity between the CO-detected and undetected galaxies, although the large uncertainties in individual estimates of the local overdensities and the small sample size of our study makes it difficult to determine whether environmental effects are responsible for the variations in \fgas and \tdep we observe in our sample.

Dynamical processes can suppress the SFR in massive galaxies below the expected rates \citep[e.g.,][]{martig09,genzel14b}, even in the presence of large molecular masses, generally referred to as morphological or gravitational quenching. In this scenario the formation of a stellar bulge stabilizes an embedded gas disk against collapse, and the stabilization can proceed from the inner regions of a galaxy outwards as the stellar surface density increases over time. These dynamical processes can accommodate relatively high gas fractions, up to $\fht \sim 0.1$, but predict long depletion times reaching 10\,Gyr or more; the molecular material is still present within the galaxies, but is prevented from forming stars.

Active, internal processes, on the other hand, such as feedback from AGN or star formation  act to suppress star formation by heating and/or removing the molecular material (see \citealt{fabian12} for a review). The energetics and driving mechanisms of these feedback processes are being actively debated, but generally predict low cold gas fractions and short depletion times.

Our observations reveal low gas fractions, $\fht \lesssim 0.1$, and short depletion times, $\tdep \lesssim 2$\,Gyr. The CO-undetected objects, in particular, show lower gas fractions and depletion times comparable to or shorter than expected for star-forming galaxies of similar mass, a conclusion that is robust even if the SFRs have been overestimated by $\sim5\times$. This demonstrates that intermediate-redshift passive galaxies are genuinely gas-poor, and the low SFRs are not the result of increasing \tdep. This suggests that dynamical processes are not acting to suppress the star formation in this sample. A similar conclusion was reached by \citet{sargent15} at slightly higher redshift, who failed to detect a single very massive galaxy at $z\sim1.5$ with $\log \Mstar/\Msol \sim 11.8$, and concluded that gravitational stabilization of gas discs is not common among the most massive high-redshift quiescent galaxies. 

Although there is some evidence that the morphological quenching scenario is consistent with the $z=0$ early-type population \citep{martig13}, we do not find evidence of it at higher redshift. There are several key differences between our sample and local massive early-types that are likely responsible for this. First, our (detected) sample galaxies contain $\approx$1--2 orders of magnitude more molecular gas than the $z=0$ population, and the upper limits on the undetected galaxies are still consistent with $>1$ order of magnitude higher \Mht. While morphological quenching can accommodate gas fractions of $\fht\sim0.1$, this may be difficult for galaxies to achieve in practice without resuming star formation. Second, our galaxies have stellar ages several Gyr younger than the $z=0$ early types (\citealt{wu18}; Chauke \etal, submitted), and so are much closer to their epoch of quenching. \citet{martig09} show that morphological quenching can act when a galaxy has already assembled a stellar spheroid and nearly depleted its gas reservoir, but is slowly re-accreting gas. It is not clear that the same should hold if the galaxy never fully depletes its gas supply, but merely continues to form stars from its existing gas reservoir.

\subsection{Connection to Local Quiescent Galaxies}

Finally, we consider the possible connection between the passive galaxies we have observed at intermediate redshifts and their counterparts in the local universe. Given the molecular masses we have observed, it is immediately clear that if these objects are indeed progenitors of local massive quiescent galaxies, they must consume, expel, or heat $\gtrsim$90--99\% of the molecular material over the next 6\,Gyr in order to match the range of \Mht seen in nearby ellipticals. This implies a much steeper drop in gas fraction compared to star-forming galaxies over the same redshift range, which decline by less than an order of magnitude over the same interval \citep[e.g.,][]{tacconi18}. 

Part of the reason for the slower decline in \fht for the star-forming population is that galaxies continue to accrete substantial amounts of gas over time from minor mergers and streams along the cosmic web \citep[e.g.,][]{keres05}. Because \tdep is observed to be much less than a Hubble time, galaxies' gas reservoirs must be replenished if they are to continue forming stars at the observed high rates. The implied gas accretion rates are $\approx100$\,\Msol/yr at $z>2$, declining by a factor of $\sim7$ from $z=2$ to $z=0.7$, and a further factor of $\sim7$ to $z=0$ \citep{scoville17}. Galaxies on and near the star-forming sequence then grow largely in equilibrium with the supply of gas made available to them \citep[e.g.,][]{dave11,lilly13,peng14}.

Massive quiescent galaxies, in contrast, are generally consistent with passive evolution over many Gyr, with growth mostly influenced by gas-poor minor mergers but relatively little further star formation. Of course, the present-day quiescent population is comprised of galaxies that quenched star formation at a variety of redshifts, with a wide variety of evolutionary histories before and after the epoch of quenching. Here, we ask whether the galaxies we have observed at $z\sim0.7$ are consistent with the population of present-day massive quiescent galaxies, assuming no external gas accretion further influences their evolution after the epoch at which we observe them.

\begin{figure}[htb]
\centering
\includegraphics[width=\columnwidth]{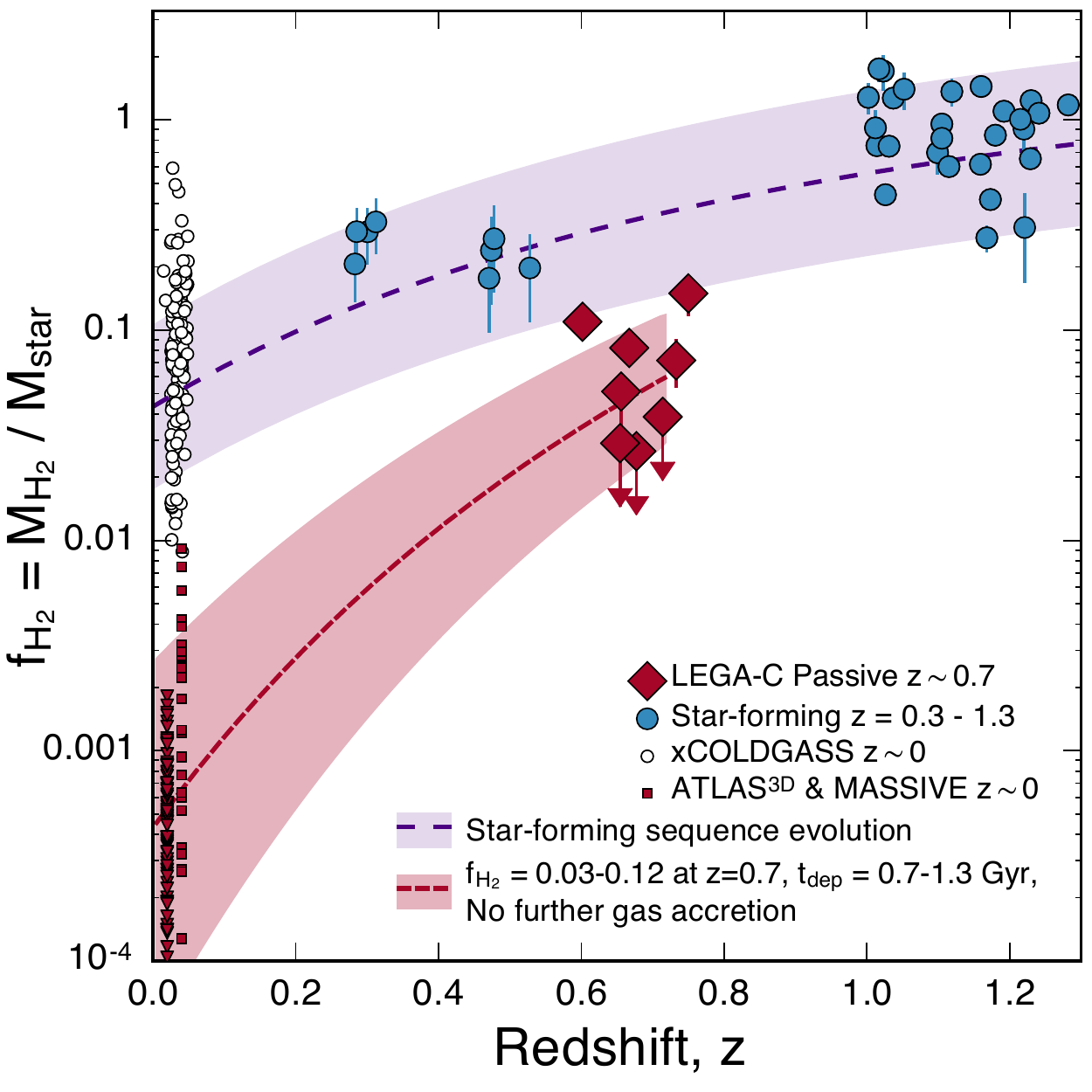}
\caption{
Evolution of \fht with redshift for galaxies on and below the star-forming sequence. Our observations are shown with red diamonds, blue circles show galaxies near the star-forming sequence at $0.3<z<1.3$ \citep{bauermeister13,tacconi13}, white circles are $z<0.05$ star-forming galaxies with $10 < \log \Mstar/\Msol < 10.7$ from xCOLDGASS \citep{saintonge17}, and small red symbols show massive ($\log \Mstar/\Msol > 10.7$) quiescent galaxies from the ATLAS$^{\mathrm{3D}}$ and MASSIVE surveys \citep{young11,davis16}. For clarity of presentation, all detections and non-detections from these last two surveys are plotted at $z=0.04$ and $z=0.02$, respectively; in reality all these galaxies are within 110\,Mpc ($z<0.025$).  The densely-dashed red line and region sketch a plausible evolutionary path for the passive galaxies we observed from $z=0.7$ to $z=0$, assuming initial starting $\fht=0.03-0.12$ and constant $\tdep=0.7-1.3$\,Gyr, and no subsequent accretion of gas. The navy dashed line and region show the evolution for galaxies with $10 < \log \Mstar/\Msol < 11$ and within a factor of 3 of the star-forming sequence from \citet{tacconi18}, where the wide range of these parameters is intended to smooth over the trajectory of individual galaxies around the star-forming sequence. The continuity of this relation and the star-forming sequence requires substantial gas accretion over cosmic time. This Figure demonstrates that the galaxies we have observed at $z\sim0.7$ can naturally reproduce the observed gas fractions of the local massive, quiescent population simply by maintaining the same (relatively short) \tdep we measure, with no need for further gas accretion or dynamical stabilization.
}
\label{fig:zevol}
\end{figure}

To address this question, we construct a closed-box toy model, in which a galaxy at $z=0.75$ with $\log \Mstar/\Msol=11$ continues to deplete its reservoir of molecular gas in a similar fashion to the galaxies we have observed. Namely, we assume a range of initial gas fractions  of $\fht=0.03-0.12$ at $z=0.7$, which decline with time as low-level star formation continues. The evolving (declining) SFRs are determined by assuming a range of depletion times similar to our observations, $\tdep=0.7-1.3$\,Gyr, that remain constant over time. We assume that a long-term average of $\approx$30\% of the mass of stars formed is returned to the ISM through stellar mass loss (\citealt{leitner11}; this is also the same assumption made by \citealt{scoville17} to infer high gas accretion rates for star-forming galaxies). Our conclusions are not sensitive to this assumption because the gas masses, and hence SFRs and additional stellar mass formed, are low. The result of this calculation is shown as the densely-dashed red line and region in Figure~\ref{fig:zevol}, and we compare to nearby  quiescent galaxies with $\log \Mstar/\Msol > 10.7$ selected from the ATLAS$^{\mathrm{3D}}$ and MASSIVE surveys \citep{young11,davis16}. 

We find that a population of intermediate-redshift galaxies with properties similar to our observed sample can very naturally reproduce the range of \fht seen in nearby passive galaxies, with no need for further gas accretion or dramatic increase in \tdep. Alternatively, if additional gas is accreted, it likely remains ionized or neutral, instead of molecular. The same conclusion does not hold for more `normal' star-forming galaxies (blue and white points, and blue dashed line and region in Figure~\ref{fig:zevol}), which show gas fractions elevated well above what would be expected in the case of negligible gas accretion. If indeed galaxies similar to those we have observed are progenitors of the local quiescent population, the molecular gas fraction must fall more rapidly with redshift than it does for star-forming galaxies in order to reconcile the observations at $z=0$ and $z\sim0.7$.

It is also worth noting that the low-level star formation in our toy model does not contradict conclusions about the ages of the stellar populations of quiescent galaxies at $z\sim0.7$ and lower redshift. \citet{wu18} compared the stellar age distributions for quiescent galaxies from \legac at intermediate redshift with the ages of similarly-massive quiescent galaxies at $z\sim0.1$ from the Sloan Digital Sky Survey. They found a typical stellar age difference of only $\sim3$\,Gyr, while the Universe aged $\approx5.5$\,Gyr -- in other words, the low-redshift quiescent galaxies are younger than would be expected in the case of pure passive evolution. \citet{wu18} surmised that either galaxies with younger stellar ages must join the quiescent population in the intervening time, or that low-level star formation in the existing quiescent galaxies must occur to lower the typical light-weighted stellar age. A true quantitative examination of this point is plagued by degeneracies, but at least qualitatively, the molecular gas fractions and depletion times observed in our sample should improve, rather than worsen, the differences in the stellar ages of quiescent objects at $z\sim0.7$ and 0.1.

Our finding that passive galaxies at $z\sim0.7$ and $z=0$ can be very naturally linked through internal processes continuing to deplete the gas reservoirs provides another independent line of evidence that preventing gas accretion and/or cooling to molecular form is key to maintaining low SFRs over many Gyr. Several proposed physical mechanisms can accomplish this, including preventing accretion of cold streams and/or satellite galaxies through the hot circumgalactic medium in massive halos \citep[e.g.,][]{birnboim03,johansson09}, and radio- or `maintenance-mode' feedback from AGN to prevent cooling \citep[e.g.,][]{croton06}. Indeed, \citet{barisic17} concluded that radio-mode feedback is likely at work in the most massive galaxies in the \legac sample, acting to maintain the observed low SFRs over several Gyr.

%%%%%%%%%%%%%%%%%%%%%%%%%%%%%%%%%%%%%%%%%%%%%%%%%%%%%%%%%%%%%%%%%%%%%%%%%%%%%%%%%%%%%
%%%%%%%%%%%%%%%%%%%%%%%%%%%%%%%%% Conclusions %%%%%%%%%%%%%%%%%%%%%%%%%%%%%%%%%%%%%%%
%%%%%%%%%%%%%%%%%%%%%%%%%%%%%%%%%%%%%%%%%%%%%%%%%%%%%%%%%%%%%%%%%%%%%%%%%%%%%%%%%%%%%
\section{Conclusions and Outlook} \label{conclusions}

In this work we have presented ALMA observations of the CO(2--1) line in a sample of 8 massive galaxies at $z\sim0.7$ selected to lie significantly below the star-forming sequence of galaxies at this redshift, drawn from the \legac spectroscopic survey. Our observations provide a first foray into understanding the molecular ISM properties of passive galaxies outside the local universe. With modest integration by ALMA, we significantly detect CO emission from four objects at $>6-8\sigma$, but do not detect the other four even after stacking, yielding upper limits on the molecular mass \Mht. At least three of the four detected galaxies show velocity gradients in the CO emission typical of rotation, and this rotation appears consistent with the stellar rotation seen in the \legac spectra.

The main focus of this work has been to understand the molecular ISM properties of these passive galaxies in the context of star-forming and quiescent galaxies at $0 < z < 1.5$. In particular, given the large investment in understanding how gas properties such as \fht and \tdep vary with other galaxy properties for the star-forming population, we examined whether galaxies further below the star-forming sequence also follow these scaling relations. We find evidence that passive galaxies at intermediate redshifts have both lower gas fractions and shorter depletion times than expected based on extrapolations from observations of the star-forming population, particularly for the objects undetected in CO emission. This suggests either a break in the scaling relations at sSFRs $\approx4-5\times$ below the star-forming sequence, or a large increase in the scatter of individual galaxy gas properties. On the other hand, a comparison to the EAGLE cosmological simulation indicates that the CO-undetected objects are in reasonable agreement with comparable simulated galaxies, while the CO-detected objects in our sample are somewhat more gas-rich than those seen in the simulation.  In both cases, the differences between observations and the predicted scaling relations are likely linked to the physical processes responsible for galaxy quenching, which are not incorporated into the purely-empirical observational scaling relations and are uncertain in the simulation.

The generally short depletion times we observe are more consistent with active feedback processes preventing star formation, rather than dynamical or gravitational stabilization against collapse of the existing gas reservoirs. We also show that a population of galaxies like those we have observed can very naturally reproduce the observed gas fractions of $z\sim0$ early-type galaxies while maintaining the depletion times we observe, with no need for additional gas accretion or increased depletion times predicted by dynamical stabilization. We see no evidence of environmental effects causing the variations in \fgas and \tdep we observe, though we stress that the small sample size of our study makes this inconclusive.

Developing a fuller understanding of the range of gas masses, fractions and depletion times in high-redshift quiescent galaxies will require observations of a significantly larger sample of objects than we have conducted here. This will allow us to determine whether passive galaxies truly show a break towards low \Mht compared to expectations or whether this is merely an artifact of limited sample size. While the dynamical information provided by CO spectroscopic observations is valuable, it is also clear that a larger sample of passive galaxies will require alternate methods of determining gas masses, such as observing long-wavelength dust continuum emission. Detecting gas fractions at the level of the stacked non-detections in CO with ALMA, for example, would require over an order of magnitude more observing time than we have invested here; reaching the same level in 870\,\um dust continuum would require about the same amount of on-source integration as our present observations. Clearly the detection of molecular gas in passive galaxies at high-redshift remains a significant challenge, even with the sensitivity of ALMA. Nevertheless, these observations present a new avenue to probe the physics of galaxy quenching and the emergence of the dichotomy between star-forming and quiescent galaxies.

\acknowledgements{
JS thanks the McDonald Observatory at the University of Texas at Austin for support through a Harlan J. Smith Fellowship.
CL is funded by a Discovery Early Career Researcher Award (DE150100618) of the Australian Research Council.
CCW acknowledges support from the National Science Foundation Astronomy and Astrophysics Fellowship grant AST-1701546.
This paper makes use of the following ALMA data: ADS/JAO.ALMA\#2016.1.00790.S. ALMA is a partnership of ESO (representing its member states), NSF (USA) and NINS (Japan), together with NRC (Canada), NSC and ASIAA (Taiwan), and KASI (Republic of Korea), in cooperation with the Republic of Chile. The Joint ALMA Observatory is operated by ESO, AUI/NRAO and NAOJ. The National Radio Astronomy Observatory is a facility of the National Science Foundation operated under cooperative agreement by Associated Universities, Inc.
Based on observations made with ESO Telescopes at the La Silla Paranal Observatory under programme ID 194-A.2005 (The \legac Public Spectroscopic Survey).
}

\facility{ALMA}

\bibliographystyle{yahapj}
%\bibliography{../../../spt_js_merge.bib}
\bibliography{legac_alma.bbl}

\end{document}